\documentclass[aps,prd,showpacs,twocolumn,groupedaddress]{revtex4}
\usepackage{graphicx}
\usepackage{amsfonts}
\usepackage{amssymb}
\usepackage{amsmath}

\begin{document}

\title{Gluon-propagator functional form  
   in the Landau gauge in SU(3) lattice QCD: \\
Yukawa-type gluon propagator and anomalous gluon spectral function
}

\author{Takumi~Iritani}
\affiliation{Department of Physics, Graduate School of Science, Kyoto University, \\
Kitashirakawa-oiwake, Sakyo, Kyoto 606-8502, Japan}
\author{Hideo~Suganuma}
\affiliation{Department of Physics, Graduate School of Science, Kyoto University, \\
Kitashirakawa-oiwake, Sakyo, Kyoto 606-8502, Japan}
\author{Hideaki~Iida}
\affiliation{The Institute of Physical and Chemical Research (RIKEN), \\
Hirosawa 2-1, Wako, Saitama 351-0198, Japan}
\date{\today}

\begin{abstract}
We study the gluon propagator $D_{\mu\nu}^{ab}(x)$ in the Landau gauge 
in SU(3) lattice QCD at $\beta$ = 5.7, 5.8, and 6.0 at the quenched level.
The effective gluon mass is estimated as $400 \sim 600$MeV 
for $r \equiv (x_\alpha x_\alpha)^{1/2} = 0.5 \sim 1.0$ fm.
Through the functional-form analysis of $D_{\mu\nu}^{ab}(x)$ obtained 
in lattice QCD, we find that the Landau-gauge gluon propagator 
$D_{\mu\mu}^{aa}(r)$
is well described by the Yukawa-type function 
$e^{-mr}/r$ with $m \simeq 600$MeV for $r = 0.1 \sim 1.0$ fm 
in the four-dimensional Euclidean space-time.
In the momentum space, the gluon propagator $\tilde D_{\mu\mu}^{aa}(p^2)$ 
with $(p^2)^{1/2}= 0.5 \sim 3$ GeV 
is found to be well approximated with a new-type propagator of $(p^2+m^2)^{-3/2}$, 
which corresponds to the four-dimensional Yukawa-type propagator.
Associated with the Yukawa-type gluon propagator, 
we derive analytical expressions for 
the zero-spatial-momentum propagator $D_0(t)$, 
the effective mass $M_{\rm eff}(t)$, and the spectral function 
$\rho(\omega)$ of the gluon field. 
The mass parameter $m$ turns out to be the effective gluon mass 
in the infrared region of $\sim$ 1fm. 
As a remarkable fact, the obtained gluon spectral function $\rho(\omega)$ 
is almost negative-definite for $\omega >m$, except for a positive $\delta$-functional peak at $\omega=m$.
\end{abstract}


\maketitle

\section{Introduction}

Quantum chromodynamics (QCD) and the gluon field were first proposed by Nambu in 1966 \cite{N66} 
just after the introduction of color degrees of freedom \cite{HN65}, 
and QCD has been established as the fundamental gauge theory of the strong interaction, 
through the explanation of asymptotic freedom \cite{GWP73},
the success of perturbative QCD for high-energy phenomena based on the parton model \cite{F69BP69,GSS07}, 
and lattice QCD calculations \cite{W74KS75,C7980,R05}.
In spite of many successes of QCD, there are still unsolved problems in the low-energy region of QCD, 
owing to its strong-coupling nature.
Indeed, the mechanism of color confinement is regarded as 
one of the most difficult important problems in theoretical physics \cite{M00}, 
and spontaneous chiral-symmetry breaking \cite{NJL61} is also a difficult issue 
in describing it quantitatively directly from QCD \cite{H91M93}. 

One of the difficulties of QCD lies on the large gap 
between the fundamental fields (quarks and gluons) and the observable particles (hadrons). 
In fact, quarks and gluons, which are the building blocks of QCD, have no physical asymptotic states,
as a result of color confinement or the gauge invariance.
This is in contrast to the ordinary perturbation theory, where 
all the observable phenomena can be directly described with 
the fundamental fields appearing in the Lagrangian.
Furthermore, the field-theoretical description of the confined particles 
is an interesting but unsolved difficult subject. 
In general, the Green function is one of the most basic quantities 
to describe the motions and the interactions of particles \cite{M99},
but most Green's functions of quarks and gluons are still unknown in the nonperturbative description.

As for the essence of nonperturbative QCD, the central issue is gluon dynamics rather than quarks \cite{C8207}.
In fact, the strong gluon interaction makes the QCD vacuum highly nontrivial, 
and color confinement and chiral symmetry breaking are realized even at the quenched level \cite{R05}. 
Then, the analysis of gluon properties is the key point to clarify the nonperturbative aspects of QCD.
In particular, the gluon propagator, {\it i.e.}, the two-point Green function is one of the most basic quantities in QCD, 
and has been investigated with much interest in various gauges, such as the Landau gauge 
\cite{M99,G78,Z91920204,S868790,SAH9701,ABP08,K09,MO87,GGKPSW87,BPS94,MMST9395,C9798,UK9899,B9900,BBLW0001,LRG02,FN04,BHLPW0407,SIMPS05,BIMPS0907,SO0607,CM0708}, 
the Coulomb gauge \cite{GOZ0304,CZ02}, and the maximally Abelian (MA) gauge \cite{AS99,K9800}, 
in the context of various aspects of QCD.
For example, an infrared-vanishing gluon propagator \cite{Z91920204,SAH9701} 
is proposed from the mathematical analysis of the Gribov horizon \cite{G78,C9798}, 
and the infrared singularity of the gluon propagator has been investigated 
in terms of color confinement \cite{M79,BBZ8191,SST95}, 
and also from the viewpoint of renormalon cancellation \cite{BB94B99,BSV01S03}.

Dynamical gluon-mass generation \cite{C8207} is also an important subject related to the infrared gluon propagation. 
In spite of massless perturbative gluons, the gluon field is conjectured to acquire a large effective mass 
as the self-energy through the self-interaction of gluons in a nonperturbative manner.
For, the glueball states, color-singlet bound states of gluons, are considered to be fairly massive 
{\it e.g.}, about 1.5GeV for the lowest $0^{++}$ and about 2GeV for the lowest $2^{++}$, 
as indicated in lattice QCD calculations \cite{R05,ISM02}. 
So far, a large effective gluon mass estimated about $0.4 \sim 0.8$GeV has been proposed in various context of QCD physics:
analytical framework based on QCD \cite{C8207}, lattice QCD calculations \cite{MO87,GGKPSW87,BPS94,MMST9395,B8283}, 
Pomeron physics \cite{HKN93}, glueball phenomenology \cite{CS83,HLW01}, 
and heavy-quark phenomenology \cite{PP80}.
Since the color ${\rm SU}(N_c)$ symmetry is unbroken in QCD, 
the effective-mass generation of gluons is quite different from the Higgs mechanism, which is a 
standard mass generation in quantum field theories.
In the electro-weak unified theory, the weak bosons, $W_\mu$ and $Z_\mu$, 
obtain quite a large mass of about 100GeV, as a result of the Higgs mechanism, 
{\it i.e.}, spontaneous breaking of the ${\rm SU(2)}\times {\rm U}(1)$ gauge symmetry.
However, the gluon mass generation is not a result of spontaneous gauge-symmetry breaking, 
but stems from much more complicated nonperturbative dynamics of gluons, 
and generally depends on the gauge choice.

The nonperturbative effects originate from the strong-coupling infrared region of QCD.
Actually, a recent lattice QCD study clarifies that 
the relevant energy scale for confinement is the infrared gluon component below 1.5GeV \cite{YS08}:
the string tension is almost unchanged even by cutting off high-momentum gluons above 1.5GeV.
Reflecting the asymptotic freedom or the running coupling $\alpha_s(\mu^2)$, 
QCD exhibits various different features according to the energy scale.
Here, we roughly classify three scale regions of QCD as 
``ultraviolet (UV)", ``infrared (IR)/intermediate (IM)", and ``deep-infrared (Deep-IR)" regions,
in terms of the length $r=(x_\mu x_\mu)^{1/2}$.
\begin{itemize}
\item
We define the UV region as $r <$ 0.1fm, 
which corresponds to the high-energy region more than a few GeV's.
In this region, perturbative QCD is approximately applicable to the reaction process, 
and the inter-quark potential is almost Coulomb-like \cite{STI04}.
\item
We define the IR/IM region as 0.1fm $\stackrel{<}{\sim} r \le$ 1fm, 
which ranges from a few hundred MeV to a few GeV in energy.  
In this region, the system is described by quark-gluon degrees of freedom 
in a nonperturbative way, which is usually substituted with some effective models \cite{GSS07}.
\item
We define the Deep-IR region as $r >$ 1fm, which is low-energy 
below $\Lambda_{\rm QCD}\sim$ 0.2GeV.
In this region, the perturbative running coupling $\alpha_s(\mu^2)$ diverges, 
and the confinement effect is extremely large, so that 
quark-gluon degrees freedom are hidden and the system is described by hadrons.
\end{itemize}

So far, the gluon propagator has been studied mainly in the Landau gauge 
both in analytic framework \cite{G78,Z91920204,SAH9701,ABP08,K09} and 
in lattice QCD \cite{MO87,GGKPSW87,BPS94,MMST9395,C9798,UK9899,B9900,BBLW0001,LRG02,FN04,BHLPW0407,SIMPS05,BIMPS0907,SO0607,CM0708}.
In the UV region, the gluon propagator takes a massless perturbative form, $1/p^2$, apart from the tensor factor.
In the IR/IM and the Deep-IR regions, the perturbative approach breaks down, 
and we need a nonperturbative approach such as lattice QCD calculations \cite{R05}.
Basic important ideas were proposed and investigated 
in the pioneering early-time lattice studies \cite{MO87,GGKPSW87,BPS94,MMST9395,C9798}, 
and recent lattice studies \cite{UK9899,B9900,BBLW0001,LRG02,FN04,BHLPW0407,SIMPS05,BIMPS0907,SO0607,CM0708} 
followed them and presented high-precision data with the main interests in 
the Deep-IR behavior by using huge-volume lattices \cite{BIMPS0907,SO0607,CM0708}.
Nevertheless, the functional form of the gluon propagator is unclear still now.

In this paper, we study the functional form of the gluon propagator 
in the Landau gauge in SU(3) lattice QCD Monte Carlo calculations, 
especially for {\it the IR/IM region} of $r=0.1 \sim 1.0$fm, 
which is considered to be relevant for the quark-hadron physics \cite{H91M93,IOS05},
and also aim to describe nonperturbative gluon properties, 
based on the obtained function form of the gluon propagator.

The organization of this paper is as follows.
In Sec.II, we give the formalism of the Landau gauge fixing and 
the gluon propagator both in continuum and in lattice QCD.
In Sec.III, we show the lattice QCD result of the gluon propagator 
in the Landau gauge both in the coordinate space and in the momentum space.
In Sec.IV, we estimate the effective gluon mass 
from the gluon propagator and the effective-mass plot in lattice QCD. 
In Sec.V, we investigate the functional form of the gluon propagator in the Landau gauge,
by analyzing the lattice QCD data. 
We show that the Landau-gauge gluon propagator is well described with the Yukawa-type function in the IR/IM region.
In Sec.VI, as the applications of the Yukawa-type gluon propagator, 
we derive analytic expressions for the zero-spatial-momentum propagator, the effective mass,
and the spectral function of the gluon field.
Section VII is devoted to summary and discussions.

\section{Formalism for gluon propagator in Landau gauge}
In this section, we review the formalism of the Landau gauge fixing
and the gluon propagator in the Euclidean space-time.

The Landau gauge is one of the most popular gauges in QCD. 
As a remarkable feature, the Landau gauge keeps Lorentz covariance and global ${\rm SU}(N_c)$ symmetry.
Owing to these symmetries and the transverse property, the color and Lorentz structure of the gluon propagator is uniquely determined.

In the Euclidean space-time formalism such as lattice QCD, 
the Landau gauge is usually defined so as to minimize the gauge-field fluctuation. 
We then expect that only the minimal fluctuation of the gluon field survives in the Landau gauge, 
and the physical essence of gluon properties can be investigated 
without suffering from large stochastic fluctuations of gauge degrees of freedom.

\subsection{Landau gauge fixing}

To begin with, let us consider the Landau gauge fixing 
and its physical meaning in Euclidean QCD.
In the SU$(N_c)$ continuum QCD, the gluon field is expressed as 
$A_\mu(x)=A_\mu^a(x)T^a \in \mathfrak{su}(N_c)$ 
with the generator $T^a (a = 1,2,\dots, N_c^2-1)$ and $A_\mu^a(x) \in \mathbb{R}$.
The Landau gauge is usually defined by the local condition 
on the gauge field as
\begin{equation}
\partial_\mu A_\mu(x) = 0.
\label{eq:localLandau}
\end{equation}
In Euclidean QCD, the Landau gauge has a global definition 
to minimize the global quantity, 
\begin{equation}
R \equiv \int d^4 x \ {\rm Tr} \{A_\mu(x) A_\mu(x)\} = \frac{1}{2} \int d^4 x A_\mu^a(x) A_\mu^a(x),
\end{equation}
by the gauge transformation.
This global definition is more strict, and 
the local condition (\ref{eq:localLandau}) is derived from the minimization of $R$.
The global quantity $R$ can be regarded as the total amount of the gauge-field fluctuation in the Euclidean space-time.
In the global definition, the Landau gauge has a clear physical interpretation that it maximally suppresses
the artificial gauge-field fluctuation relating to the gauge degrees of freedom.

In lattice QCD, the theory is formulated on the discretized space-time \cite{R05}.
The QCD action is constructed from the link-variable $U_\mu(x) \in {\rm SU}
(N_c)$, instead of the gauge field $A_\mu(x) \in \mathfrak{su}(N_c)$.
The link-variable is defined as $U_\mu(x) \equiv e^{iagA_\mu(x)}$, with the 
lattice spacing $a$ and the gauge coupling constant $g$. The gauge transformation 
of the link-variable is given by
\begin{equation}
U_\mu(x) \rightarrow \Omega(x)  U_\mu(x) \Omega^\dagger(x+ \hat{\mu}),
\end{equation}
with the gauge function $\Omega(x) \in {\rm SU}(N_c)$.

In lattice QCD, the Landau gauge fixing is also expressed in terms of the link-variable: 
the Landau gauge is defined by the maximization of 
\begin{equation}
R_{\rm latt} = \sum_x \sum_\mu {\rm Re} \ {\rm Tr} \ U_\mu(x),
\label{eq:Rlatt}
\end{equation}
by the gauge transformation of the link-variable.

For small $a$, using the expansion  
\begin{equation}
U_\mu(x) = 1 + iag A_\mu(x) - \frac{1}{2} a^2 g^2 A_\mu^2(x) + O(a^3)
\label{eq:linkVariableExpansion}
\end{equation}
in terms of lattice spacing $a$, 
$R_{\rm latt}$ is expressed as
\begin{equation}
R_{\rm latt} = - \frac{a^2 g^2}{4} \sum_x A_\mu^a (x) A_\mu^a(x)
+ O(a^4) + {\rm const.} 
\end{equation}
Therefore, the maximization of $R_{\rm latt}$ 
corresponds to the minimization of the gauge-field fluctuation as well as the continuum theory.
This minimization of the gluon-field fluctuation in turn 
justifies the expansion in Eq.(\ref{eq:linkVariableExpansion}).

\subsection{Gluon propagator in lattice QCD}

In this subsection, we formulate the gluon propagator in the Landau gauge.
To begin with, we extract the gluon field from the gauge-fixed link-variable,
which is obtained by the gauge transformation of the link-variable 
to maximize $R_{\rm latt}$.
Using the expansion (\ref{eq:linkVariableExpansion}), 
we define the bare gluon field $A^{\rm bare}_\mu(x)$ as
\begin{equation}
A^{\rm bare}_\mu(x) \equiv \frac{1}{2iag} 
\left[ U_\mu(x) - U_\mu^\dagger(x) \right] 
- \frac{1}{2iagN_c} {\rm Tr} \left[ U_\mu(x) - U_\mu^\dagger(x) \right],
\label{eq:gluonfield}
\end{equation}
where the second term is added to make $A^{\rm bare}_\mu$ traceless.
The renormalized gluon field $A^{\rm ren}_\mu(x)$ is obtained by multiplying 
a real renormalization factor $Z_3^{-1/2}$ as 
\begin{equation}
A^{\rm ren}_\mu(x) \equiv Z_3^{-1/2} A^{\rm bare}_\mu(x).
\label{eq:gluonfield}
\end{equation}
We abbreviate $A^{\rm ren}_\mu(x)$ as $A_\mu(x)$ hereafter.
The gluon field $A_\mu(x)$ defined above is traceless and hermite, 
and is expressed as $A_\mu(x)=A_\mu^a(x)T^a \in \mathfrak{su}(N_c)$ 
with $A_\mu^a(x) \in \mathbb{R}$.
In the Landau gauge which maximizes $R_{\rm latt}$, 
the gluon field $A_\mu(x)$ satisfies the local condition, 
\begin{eqnarray}
\partial_\mu A_\mu (x)=0,
\label{eq:localLandauL}
\end{eqnarray}
with the forward or backward derivative $\partial_\mu$ on the lattice.

The gluon propagator $D_{\mu\nu}^{ab}(x)$ is defined by the two-point function as  
\begin{equation}
D_{\mu\nu}^{ab}(x,y) \equiv \langle A_\mu^a(x) A_\nu^b(y) \rangle=D_{\mu\nu}^{ab}(x-y).
\end{equation}
Note that the time-ordered product is unnecessary in the Euclidean metric, 
and the translational invariance of the vacuum leads to the $(x-y)$-dependence.
From Eq.(\ref{eq:localLandauL}), $D_{\mu\nu}^{ab}(x-y)$ satisfies the transverse property, 
\begin{equation}
\partial^x_\mu D_{\mu\nu}^{ab}(x-y) =\partial^y_\nu D_{\mu\nu}^{ab}(x-y) =0.
\end{equation}

Next, we consider the gluon propagator $\tilde{D}_{\mu\nu}^{ab}(p)$ in the momentum space, 
which is defined by the Fourier transformation of the coordinate-space propagator as 
\begin{equation}
\tilde{D}_{\mu\nu}^{ab}(p) \equiv \int d^4 x \ e^{ip\cdot x} D_{\mu\nu}^{ab}(x).
\end{equation}
On the $L_1 \times L_2 \times L_3 \times L_4$ lattice,
this Fourier transformation is discretized, 
and the momentum-space gluon propagator is expressed as
\begin{equation}
\tilde{D}_{\mu\nu}^{ab}(p) = \sum_x e^{i\hat{p} x}D_{\mu\nu}^{ab}(x).
\end{equation}
Here, the discretized momentum $\hat{p}_\mu$ 
and the continuum momentum $p_\mu$ are defined as \cite{R05,M99,UK9899,BBLW0001}
\begin{equation}
\hat{p}_\mu \equiv \frac{2\pi n_\mu}{aL_\mu}, \quad
p_\mu \equiv \frac{2}{a} \sin \left(\frac{\hat{p}_\mu a}{2}\right)
= \frac{2}{a} \sin \left(\frac{\pi n_\mu}{L_\mu}\right),
\end{equation}
with $n_\mu = 0, 1, 2, ... , L_\mu-1$.

In the Landau gauge, the color and tensor structure of the propagator is uniquely determined as
\begin{equation}
\tilde{D}_{\mu\nu}^{ab}(p) 
= \tilde{D}(p^2) \delta^{ab} \left( \delta_{\mu\nu} - \frac{p_\mu p_\nu}{p^2}
\right),
\label{eq:GPmom}
\end{equation}
from the ${\rm SU}(N_c)$ global symmetry, the Lorentz symmetry, and the transverse property, 
\begin{equation}
p_\mu\tilde{D}_{\mu\nu}^{ab}(p) =p_\nu\tilde{D}_{\mu\nu}^{ab}(p) =0.
\end{equation}
Therefore, we only have to consider the scalar factor of the gluon propagator, $\tilde D(p^2)$, 
which is a function of the continuum-momentum squared $p^2=p_\alpha p_\alpha$.

In the coordinate space, we investigate the scalar combination of the gluon propagator 
\begin{equation}
D(r) \equiv \frac{1}{3 (N_c^2-1) } D_{\mu\mu}^{aa}(x)
     = \frac{1}{3 (N_c^2-1) } 
\langle A_\mu^a(x)A_\mu^a(0)\rangle,
\label{eq:defPropagator}
\end{equation}
as a function of the four-dimensional Euclidean distance,
\begin{equation}
r \equiv |x| \equiv (x_\mu x_\mu)^{1/2}.
\end{equation}
Here, the denominator factor $3$ in Eq.(\ref{eq:defPropagator}) has been introduced 
considering the tensor factor 
$\delta_{\mu\mu} - p_\mu p_\mu/p^2 = 3$. 

The scalar factor $\tilde{D}(p^2)$ in Eq.(\ref{eq:GPmom}) is expressed 
by the Fourier transformation of $D(r)$ as 
\begin{equation}
\tilde{D}(p^2) = \frac{1}{3 (N_c^2-1)} \tilde{D}_{\mu\mu}^{aa}(p) 
= \sum_x e^{i\hat{p} \cdot x} D(r),
\label{eq:defPropagatorP}
\end{equation}
from which one can prove that $D(r)$ depends only on $r$ near the continuum limit.
In this paper, we call $D(r)$ and $\tilde{D}(p^2)$ ``scalar-type propagator".

\section{Lattice QCD result for gluon propagator}

We perform ${\rm SU}(3)$ lattice QCD Monte Carlo calculations at the quenched level
using the standard plaquette action. 
Here, we adopt three different lattices with the lattice parameter 
$\beta \equiv 2N_c/g^2$=5.7, 5.8, and 6.0.
The used lattice size is $16^3 \times 32$, $20^3 \times 32$, and $32^4$
at $\beta$ = 5.7, 5.8, and 6.0, respectively.
In this lattice calculation, we mainly use $\beta=6.0$.

The lattice spacing $a$ is found to be $a = 0.186, 0.152$, and $0.104$fm, 
at $\beta$ = 5.7, 5.8, and 6.0, respectively, 
when the scale is determined so as to reproduce the string tension as 
$\sqrt{\sigma} = 427$MeV from the static Q$\bar {\rm Q}$ potential \cite{STI04}. 

The number of used gauge configurations is 50, 40 and 30 for $\beta$ =5.7, 5.8, and 6.0, respectively.
The gauge configurations are picked up every 1,000 sweeps after a thermalization of 20,000 sweeps.
We summarize the parameter and calculation conditions in Table~\ref{tab:latParam}.

\begin{table}[h]
\begin{center}
\caption{\label{tab:latParam}
The lattice parameter $\beta$, lattice size, and 
the gauge-configuration number $N_{\rm conf}$. 
The corresponding lattice spacing $a$ and 
the lattice volume in the physical unit are added.
The lattice spacing $a$ is determined so as to reproduce 
the string tension $\sqrt{\sigma} = 427$MeV.
}
\begin{tabular}{ccccc}
\hline\hline
~~$\beta$~~ & ~~Lattice size~~ & ~~$a$ [fm]~~ 
  & ~~Volume [${\rm fm}^4$]~~ & ~~$N_{\rm conf}$~~ \\
\hline
5.7 & $16^3\times 32$ & 0.186 & $2.976^3 \times 5.952$ & 50 \\
5.8 & $20^3\times 32$ & 0.152 & $3.040^3 \times 4.864$ & 40 \\
6.0 & $32^3\times 32$ & 0.104 & $3.328^3 \times 3.328$ & 30 \\
\hline\hline
\end{tabular}
\end{center}
\end{table}

Here, we briefly explain the actual procedure of the gluon-propagator calculation.
For each gauge configuration, 
we perform the Landau gauge fixing by the gauge transformation to 
maximize $R_{\rm lat}$ defined in Eq.(\ref{eq:Rlatt}), 
and obtain the gluon field $A_\mu(x)$ 
defined in Eq.(\ref{eq:gluonfield}) from the gauge-fixed link-variable.
Then, we construct the scalar-type gluon propagator 
$D(r)$ from the two-point function of the gluon field, as shown in Eq.(\ref{eq:defPropagator}).
Finally, we calculate the momentum-space gluon propagator $\tilde{D}(p^2)$ 
using the discrete Fourier transformation, as shown in Eq.(\ref{eq:defPropagatorP}).

As for the overall renormalization factor $Z_3$, 
the gluon field $A_\mu(x)$ is renormalized so as to make  
the gluon propagator $\tilde D(p^2)$ 
coincide with the tree-level propagator $1/p^2$ at $\mu$ \cite{BBLW0001}, {\it i.e.}, 
\begin{equation}
\tilde{D}(p^2)\Big|_{p^2=\mu^2} = \frac{1}{\mu^2}.
\label{RenormCond}
\end{equation}

In the calculation of the gluon propagator, 
we take the advantage of the translational symmetry to improve statistics,
and we adopt the jackknife method to estimate the statistical error.

\begin{figure}[h]
\begin{center}
\includegraphics[scale=1.1]{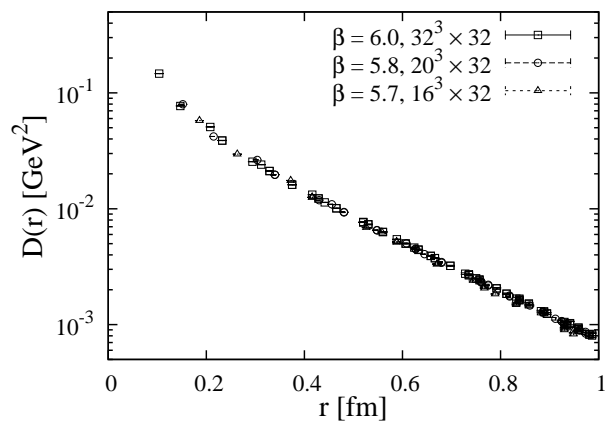}
\caption{\label{fig:latResCor}
Lattice QCD results of the scalar-type gluon propagator
$D(r) \equiv \sum_{a=1}^8\sum_{\mu=1}^4\langle A_\mu^a(x)A_\mu^a(0)\rangle/24$
as the function of the four-dimensional Euclidean distance $r \equiv (x_\alpha x_\alpha)^{1/2}$ 
in the Landau gauge at $\beta$ = 5.7, 5.8, and 6.0.
}
\end{center}
\end{figure}

\begin{figure}[h]
\begin{center}
\includegraphics[scale=1.1]{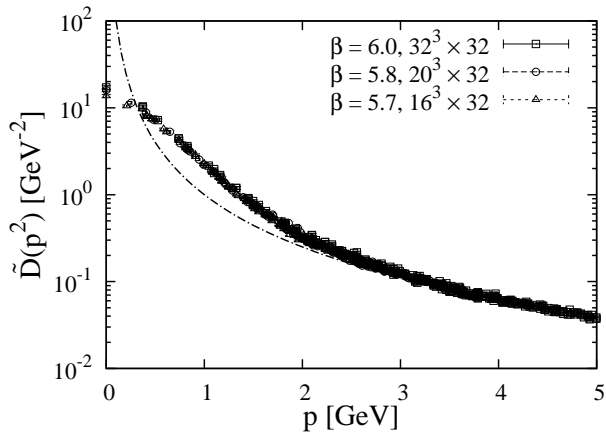}
\caption{\label{fig:latResMom}
Lattice QCD results of the scalar-type gluon propagator 
$\tilde{D}(p^2)=\sum_x e^{i\hat{p}\cdot x}D(r)$ 
plotted against $p \equiv (p_\mu p_\mu)^{1/2}$ with the momentum $p_\mu = \frac{2}{a} \sin (\frac{\pi n_\mu}{L_\mu})$, 
in the Landau gauge at $\beta$ = 5.7, 5.8, and 6.0. 
We renormalize the propagator to satisfy the renormalize condition $D(p^2)|_{p^2=\mu^2}=1/\mu^2$ at the scale
$\mu=4{\rm GeV}$. The dash-dotted line denotes the tree-level massless propagator, $1/p^2$.
}
\end{center}
\end{figure}

Figure~\ref{fig:latResCor} and \ref{fig:latResMom} 
show the lattice QCD results at $\beta$=5.7, 5.8, and 6.0 
for the scalar-type gluon propagator $D(r)\equiv D_{\mu\mu}^{aa}(x)/24$ and 
$\tilde{D}(p^2) \equiv \tilde D_{\mu\mu}^{aa}(p)/24$, respectively. 
These figures include not only on-axis data but also off-axis data.
In these figures, the statistical error is rather small in the depicted region, 
and the statistical error bars are hidden in the symbols. 
Here, we choose the renormalization scale at $\mu=4{\rm GeV}$ for $\beta=6.0$ 
\cite{BBLW0001}, and make corresponding rescaling for $\beta$=5.7 and 5.8.
We thus obtain $D(r)$ as a single-valued function of 
the four-dimensional Euclidean distance $r \equiv (x_\alpha x_\alpha)^{1/2}$.

The lattice result of $\tilde{D}(p^2)$ is almost 
a single-valued function of the magnitude of the continuum momentum, $p \equiv (p_\alpha p_\alpha)^{1/2}$.
We confirm that our lattice QCD result of $\tilde D(p^2)$ is 
consistent with that obtained in the previous lattice studies \cite{UK9899,BBLW0001,FN04},
although recent huge-volume lattice studies \cite{BIMPS0907,SO0607,CM0708} 
indicate a suppression of the gluon propagator in the Deep-IR region ($p < 0.5$GeV),
compared with the smaller lattice result.
The lattice data of $\tilde D(p^2)$ seem to be consistent with  
the tree-level massless propagator $\tilde D_{\rm tree}(p^2) =1/p^2$ for large $p^2$, 
but their behaviors are largely different in the low-energy region below a few GeVs.

\section{\label{sec:massAnalysis}Effective gluon mass}

In this section, we investigate the effective gluon mass  
in the Landau gauge using the gluonic correlation obtained in lattice QCD. 
We derive the massive-vector propagator in coordinate space, 
and estimate the effective gluon mass, 
by comparing the lattice gluon propagator with the massive propagator. 
Also, we investigate the effective-mass plot of gluons 
obtained from the zero-spatial-momentum propagator in lattice QCD.

\subsection{Comparison with massive propagator}

We first derive the free massive-vector propagator form in coordinate space 
as a useful guide to analyze the gluon propagator in lattice QCD.
Here, we use the Stueckerberg form Lagrangian \cite{IZ80} 
in the Euclidean metric, 
\begin{equation}
\mathcal{L} = \frac{1}{4} \left( \partial_\mu A_\nu^a - \partial_\nu A_\mu^a \right)^2
+ \frac{1}{2} m^2 A_\mu^a A_\mu^a - \frac{1}{2\alpha} 
\left( \partial_\mu A_\mu^a \right)^2,
\end{equation}
where the parameter $\alpha = 0$ corresponds to the Landau gauge.
The propagator of the massive vector field $A_\mu^a$ is derived from the Lagrangian as
\begin{equation}
\tilde{D}_{\mu\nu}^{ab}(p) 
= \frac{1}{p^2+m^2} \delta^{ab}
\left\{ \delta_{\mu\nu} - \frac{(1+ \alpha) p_\mu p_\nu}{p^2 - \alpha m^2} \right\}.
\label{eq:StuPro}
\end{equation}
Taking $\alpha = 0$, 
the massive-vector propagator in the Landau gauge is obtained as 
\begin{equation}
\tilde{D}_{\mu\nu}^{ab} (p) 
= \frac{1}{p^2 + m^2} \delta^{ab} \left(\delta_{\mu\nu} - \frac{p_\mu p_\nu}{p^2}
\right).
\end{equation}
This propagator satisfies the transverse property of the Landau gauge, 
$p^\mu \tilde{D}_{\mu\nu}^{ab}(p) = p^\nu \tilde{D}_{\mu\nu}^{ab}(p)$ = 0,
which corresponds to the condition, $\partial_\mu A_\mu^a(x) = 0$.

In this case, the scalar-type propagator $\tilde{D}(p^2)$ reads 
\begin{equation}
\tilde{D}(p^2) = \frac{1}{24}\tilde{D}_{\mu\mu}^{aa}(p)=\frac{1}{p^2+m^2},
\end{equation}
and its Fourier transformation gives the scalar-type propagator $D(r)$
in the coordinate space as \cite{AS99}  
\begin{eqnarray}
D(r) = \int \frac{d^4p}{(2\pi)^4} e^{-ip\cdot x}\tilde{D}(p^2)  
= \frac{1}{4\pi^2} \frac{m}{r} K_1(mr),
\label{eq:Dmassive}
\end{eqnarray}
where $K_1(mr)$ is the modified Bessel function. 
The derivation of this formula is shown in Appendix~A.
For large $r$, $K_1(mr)$ behaves asymptotically as 
\begin{equation}
K_1(mr) \simeq \sqrt{\frac{\pi}{2mr}} e^{-mr},
\label{eq:k1asymptotic}
\end{equation}
and therefore the massive propagator behaves as
$D(r) \sim r^{-3/2}e^{-mr}$.

Now, we compare the scalar-type gluon propagator $D(r)$ obtained in lattice QCD
with the massive propagator, 
and estimate the effective gluon mass through the fit-analysis.
Considering the functional form of the massive propagator as shown in Eq.(\ref{eq:Dmassive}),
we here adopt the fit-function defined by 
\begin{equation}
D_{\rm mass}(r) = A \frac{m}{r} K_1(mr),
\label{eq:DmassiveFit}
\end{equation}
with the mass parameter $m$ and a dimensionless parameter $A$.

\begin{figure}[h]
\begin{center}
\includegraphics[scale=1.1]{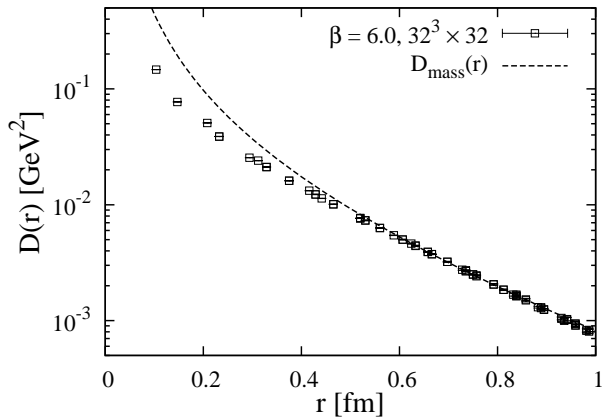}
\caption{\label{fig:DmassiveFitting}
A typical example of the fit-analysis of the lattice gluon propagator $D(r)$ 
with the fit-function $D_{\rm mass}(r)$ of the massive-vector propagator 
denoted by the dashed line, 
The fit is done for the lattice data at $\beta = 6.0$ 
in the fit-range of $r = 0.6 \sim 1.0$fm. 
}
\end{center}
\vspace{-0.18cm}
\end{figure}

For example, we show in Fig.\ref{fig:DmassiveFitting} 
the fit result of the lattice data $D(r)$ at $\beta = 6.0$ with $D_{\rm mass}(r)$ 
in the fit-range of $r = 0.6 \sim 1.0$fm. In this range, this fit seems well, 
and the effective mass $m$ is estimated to be about 500MeV from this fit. 
However, Fig.\ref{fig:DmassiveFitting} shows that the lattice gluon propagator $D(r)$ 
cannot be described with $D_{\rm mass}(r)$ in the whole region of $r = 0.1 \sim 1.0$fm.

We investigate the fit-analysis for the lattice gluon propagator $D(r)$ with $D_{\rm mass}(r)$ 
for several fit-ranges, and the results are summarized in Table~\ref{tab:DmassiveFitSummary}.
From this fit-analysis in the Landau gauge, the effective gluon mass is estimated as 
$m = 400 \sim 600$MeV in the infrared region of $r = 0.5 \sim 1.0$fm, although 
there is a significant $r$-dependence of $m$, {\it i.e.}, $m$ is small at short distances.

\begin{table}[h]
\begin{center}
\caption{\label{tab:DmassiveFitSummary}
The fit result for the scalar-type gluon propagator $D(r)$ 
obtained in lattice QCD at $\beta$ = 5.7, 5.8, and 6.0.
The fit function is the massive-vector propagator $D_{\rm mass}(r) = A mr^{-1} K_1(mr)$.
The best fit parameters $(m, A)$ are listed together with the fit range and $\chi^2/N_{\rm df}$.
The fit data with large $\chi^2/N_{\rm df}$ are omitted.
}
\begin{tabular}{lclll}
\hline\hline
$\beta$~~ & ~~fit range [fm]~~ & $m$ [GeV]~~~~ & ~~~~$A$~~~~~~~~ & ~$\chi^2/N_{\rm df}$~~~ \\
\hline
6.0 & $0.5 \sim 0.7$ & 0.405(7) & 0.094(2) & 2.07847 \\
    & $0.6 \sim 0.8$ & 0.475(6) & 0.112(2) & 0.588604 \\
    & $0.7 \sim 0.9$ & 0.551(8) & 0.140(4) & 0.454586 \\
    & $0.8 \sim 1.0$ & 0.582(10)& 0.158(6) & 0.340416 \\
    & $0.6 \sim 1.0$ & 0.517(5) & 0.125(2) & 1.67541  \\
5.8 & $0.6 \sim 0.8$ & 0.502(10)& 0.118(4) & 1.68858 \\
    & $0.7 \sim 0.9$ & 0.549(15)& 0.138(7) & 0.773316 \\
    & $0.8 \sim 1.0$ & 0.576(12)& 0.151(7) & 0.414934 \\
    & $0.6 \sim 1.0$ & 0.526(6) & 0.126(2) & 0.576734 \\
5.7 & $0.7 \sim 0.9$ & 0.626(48)& 0.162(26)& 3.76344 \\
    & $0.8 \sim 1.0$ & 0.618(32)& 0.159(19)& 2.26721 \\
\hline\hline
\end{tabular}
\end{center}
\end{table}

\subsection{\label{sec:effmass}Effective-mass plot of gluons}

In the previous subsection, we estimate the effective gluon mass 
from the fit-analysis for the lattice gluon propagator.
Now, we estimate the effective gluon mass from  
the effective-mass analysis with zero-spatial-momentum propagator $D_0(t)$ 
in the Landau gauge \cite{M99,MO87,GGKPSW87,BPS94,MMST9395}.
This method is often used for hadrons as a standard mass measurement in lattice QCD \cite{R05}. 
For the simple notation, we use the lattice unit of $a=1$ in this subsection.

We define the zero-spatial-momentum propagator $D_0(t)$ of gluons as
\begin{eqnarray}
D_0(t) \equiv \frac{1}{24} \sum_{\vec{x}} 
\langle A_\mu^a(\vec{x},t) A_\mu^a(\vec{0},0)\rangle
= \sum_{\vec{x}} D(r),~~~
\label{eq:ZMPcorrL}
\end{eqnarray}
where the total spatial momentum is projected to be zero.
Using the translational invariance, $D_0(t)$ can be rewritten as 
the wall-to-wall correlator, 
\begin{equation}
D_0(t) \equiv \frac{1}{24 \sum_{\vec{y}}1}
\sum_a \sum_\mu 
\langle\Big\{ \sum_{\vec{x}} A_\mu^a(\vec{x},t) \Big\}\Big\{ \sum_{\vec{y}} A_\mu^a (\vec{y},0)\Big\}\rangle.
\end{equation}
In the actual lattice QCD calculation, we adopt the wall-to-wall correlator 
to improve statistics with an easy task.

The effective mass of gluons is defined by 
\begin{equation}
M_{\rm eff}(t) = \ln \{D_0(t)/D_0(t+1)\},
\label{eq:EMP}
\end{equation}
in the case of large temporal lattice size.
In the numerical analysis, we take account of the temporal periodicity 
used in lattice calculations. 
On the lattice with the temporal size $N_t(=L_4)$,  
$D_0(t)$ is expected to behave as
\begin{eqnarray}
D_0(t) \propto e^{-mt} + e^{-m(N_t -t)} 
\propto \cosh \big\{ m \big( \frac{N_t}{2} - t\big)\big\},
\end{eqnarray}
and therefore we define the effective mass $M_{\rm eff}(t)$ by
\begin{equation}
\frac{D_0(t+1)}{D_0(t)} = \frac{\cosh\left[ M_{\rm eff}(t) 
\left( N_t/2 - (t+1)\right)\right]}
{\cosh\left[ M_{\rm eff}(t) \left( N_t/2 - t\right) \right]},
\label{eq:EMcosh}
\end{equation}
which is reduced to Eq.(\ref{eq:EMP}) in the large $N_t$ limit.

\begin{figure}[h]
\begin{center}
\includegraphics[scale=1.1]{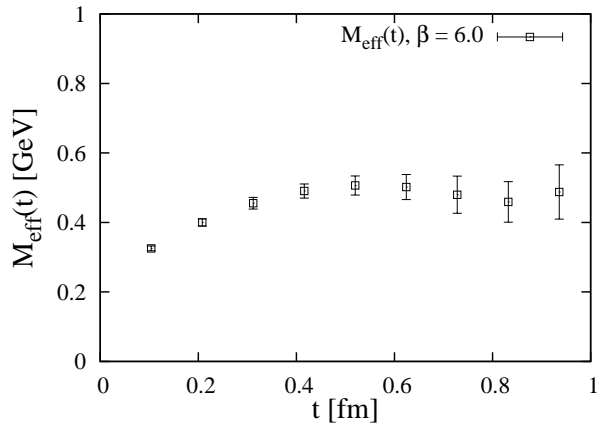}
\caption{\label{fig:effMass}
The effective mass $M_{\rm eff}(t)$ 
of gluons in the Landau gauge 
in lattice QCD at $\beta = 6.0$, {\it i.e.}, $a = 0.104$fm.
}
\end{center}
\end{figure}

Figure~\ref{fig:effMass} shows the plot of 
the effective mass $M_{\rm eff}(t)$ of gluons 
in lattice QCD at $\beta =6.0$.
The effective mass $M_{\rm eff}(t)$ of gluons 
is an increasing function for small $t$, 
and approximately constant for $t = 0.4 \sim 1.0$fm, and 
the effective gluon mass is estimated to be about 500MeV 
from the value of $M_{\rm eff}(t)$ in the range of $t = 0.4 \sim 1.0$fm.
This tendency and the estimated value are consistent with 
the previous results \cite{M99,MO87,GGKPSW87,BPS94,MMST9395} 
and those obtained from the analysis of the gluon propagator in the previous subsection.

Note here that the effective mass $M_{\rm eff}(t)$ 
of gluons exhibits an increasing behavior, which is unusual \cite{M99,MO87,GGKPSW87,BPS94,MMST9395}.
In the usual hadron-mass calculation, the effective mass 
is always a decreasing function of $t$, 
due to the positive contribution from physical excited states at the short distance.
The mathematical proof of this statement is as follows:
The zero-spatial-momentum propagator $G(t)$ is expressed in the Lehmann-K\"allen representation as 
\begin{eqnarray}
G(t)=\sum_i c_ie^{-m_it},
\end{eqnarray}
with the spectral weight $c_i$.
In the continuum formalism, the effective mass is expressed as  
$M(t)\equiv -\frac{d}{dt}\ln G(t)$ and satisfies 
\begin{eqnarray}
&&\frac{d}{dt}M(t)=-\frac{d^2}{dt^2}\ln\big( \sum_i c_ie^{-m_it}\big)
\label{eq:dMdt}
\\ \nonumber
&=&
-\frac{( \sum_i c_ie^{-m_it})( \sum_i c_i m_i^2e^{-m_it})-( \sum_i c_im_i e^{-m_it})^2
}{( \sum_i c_ie^{-m_it})^2}.
\end{eqnarray}
If all the spectral weights are non-negative as $c_i \ge 0$, 
this is always non-positive, {\it i.e.}, $\frac{d}{dt} M(t) \le 0$, 
due to the Cauchy-Schwartz inequality \cite{M99}, so that the effective mass 
$M(t)$ is generally a decreasing function in all the region of $t$.

However, the effective mass $M_{\rm eff}(t)$ of the gluon exhibits an anomalous increasing behavior, 
and this fact in turn indicates that the gluon spectral function is not positive-definite 
\cite{M99,MO87,GGKPSW87,BPS94,MMST9395} due to its unphysical nature.
Note also that, from Eq.(\ref{eq:dMdt}), we can formally obtain $\frac{d}{dt}M(t) \le 0$, 
if all the spectral weights are non-positive as $c_i \le 0$. 
In fact, the increasing property of $M(t)$, {\it i.e.}, $\frac{d}{dt}M(t) > 0$, can be realized, 
only when there is some suitable coexistence of positive and negative values in the spectral weights $c_i$.

Here, we summarize the lattice QCD result of 
the effective gluon mass in the Landau gauge.
The effective gluon mass exhibits a significant scale-dependence, 
and it takes a small value at short distances.
Quantitatively, the effective gluon mass is estimated 
to be about $400 \sim 600$MeV 
in the infrared region $\sim 1.0$fm.
This value seems consistent with the gluon mass suggested by Cornwall \cite{C8207}.

\section{\label{sec:functionalforms} Functional form of the gluon propagator in the Landau gauge}

In this section, we study the functional form of the gluon propagator in the Landau gauge in SU(3) lattice QCD.
In the high-energy region, we already know the applicability of perturbative QCD, where gluons are massless. 
The perturbative gluon propagator is simply described with $1/p^2$ in the covariant gauge, 
similar to the photon propagator in QED.
In the infrared region, however, the gluon is expected to acquire a large effective mass 
due to nonperturbative QCD effects, as was also indicated in the previous section. 
In fact, the functional form of the gluon propagator can be changed according to the scale.

So far, the functional form of the gluon propagator has been studied both 
in analytical framework \cite{G78,Z91920204,SAH9701,M79} and in lattice QCD \cite{BPS94,MMST9395,UK9899,LRG02}.
In the UV region, the lattice QCD studies have shown the perturbative behavior of the gluon propagator. 
In the Deep-IR region, the gluon propagator and its behavior have been investigated with the theoretical interest of  
``infrared vanishing" behavior in the context of the Gribov horizon, 
``dipole (1/$p^4$)-singularity" in the relation to color confinement, and so on.
Here, we aim to determine the functional form of the gluon propagator 
in the infrared/intermediate region of $r \equiv (x_\alpha x_\alpha)^{1/2} = 0.1 \sim 1.0$fm, 
which is the relevant scale of quark-hadron physics.

\subsection{Functional form candidates}

In the previous section, we compare the gluon propagator 
with the massive propagator $D_{\rm mass}(r)$, 
which corresponds to $(p^2+m^2)^{-1}$ in the momentum space.
However, as shown in Fig.\ref{fig:DmassiveFitting}, it is difficult 
to reproduce with $D_{\rm mass}(r)$ the lattice result in the whole region of $r= 0.1 \sim 1.0$fm. 
In Fig.\ref{fig:DmassiveFitting}, we choose $D_{\rm mass}(r)$ appropriate for a large-$r$ region, 
but it gives too large value than the lattice result in the small $r$ region.
In fact, at the short distance, the gluon propagator shows 
a larger reduction than the simple massive propagator.
Then, one may think of a larger mass at shorter distance in the form of $(p^2+m^2)^{-1}$. 
But, a smaller effective mass is obtained at the short distance, as was shown in the previous section.
This indicates that the functional form of the gluon propagator itself is largely changed from 
the simple massive propagator $D_{\rm mass}(r)$.

As the candidate form of the gluon propagator in the momentum space, 
we here consider $(p^2+m^2)^{-3/2}$ and $(p^2+m^2)^{-2}$, 
as well as the massive propagator, $(p^2+m^2)^{-1}$. 
Both of two candidates qualitatively satisfy the above-mentioned behavior, 
{\it i.e.}, larger reduction at the short distance.

For the actual analysis of the lattice QCD result, we mainly consider 
the gluon propagator in the coordinate space instead of the momentum space,
since the coordinate-space variable is more directly obtained in lattice QCD.

\subsubsection{Yukawa-type propagator}

First, we consider $(p^2+m^2)^{-3/2}$ type propagator. 
In the coordinate space, this corresponds to the Yukawa-type function, 
since the Fourier transformation in four-dimension Euclidean space is given by
\begin{equation}
\int \frac{d^4p}{(2\pi)^4} e^{ip\cdot x} 
\frac{1}{(p^2+m^2)^{3/2}} = \frac{1}{4\pi^2} \frac{1}{r} e^{-mr}.
\label{eq:YukawaFourier}
\end{equation}
We show the derivation of this formula in Appendix~A.

Then, we call this form ``Yukawa-type function" or ``Yukawa-type propagator". 
Usually, the Yukawa-type function is obtained by 
the three-dimensional Fourier transformation of $(p^2+m^2)^{-1}$.
It is notable that this Fourier transformation (\ref{eq:YukawaFourier}) 
is calculated in the four-dimensional Euclidean space-time, 
and therefore the momentum-space function 
takes an unfamiliar form as $(p^2+m^2)^{-3/2}$.

For the analysis of the gluon propagator, 
we introduce the definite form of 
the Yukawa-type propagator as 
\begin{equation}
D_{\rm Yukawa}(r) = A \frac{m}{r} e^{-mr},
\label{eq:Dyukawa}
\end{equation}
with a ``mass" parameter $m$ and a dimensionless parameter $A$.
With this form, we analyze the lattice QCD result of 
the gluon propagator in the coordinate space.

The Fourier transformation of $D_{\rm Yukawa}(r)$ is given by
\begin{equation}
\tilde{D}_{\rm Yukawa}(p^2) \equiv 
\int d^4 x e^{ip\cdot x} D_{\rm Yukawa}(r) 
= \frac{4\pi^2 Am}{(p^2+m^2)^{3/2}}.
\label{eq:DyukawaP}
\end{equation}
This functional form will be used for the analysis of 
the gluon propagator in the momentum-space in Sec.V-C. 

\subsubsection{Dipole-type propagator}

First, we consider $(p^2+m^2)^{-2}$ type function, which we call ``dipole-type".
In the coordinate space, this function corresponds to 
the modified Bessel function $K_0(mr)$, because of the four-dimensional Fourier transformation, 
\begin{equation}
\int \frac{d^4p}{(2\pi)^4} e^{ip\cdot x} \frac{1}{(p^2+m^2)^2} = \frac{1}{8\pi^2}K_0(mr).
\end{equation}
The derivation of this formula is shown in Appendix~A.

We define the dipole-type propagator $D_{\rm dipole}(r)$ as
\begin{equation}
D_{\rm dipole}(r) = A m^2 K_0(mr),
\end{equation}
with a ``mass" parameter $m$ and a dimensionless parameter $A$.
From the asymptotic form of the modified Bessel function,
$D_{\rm dipole}(r)$ behaves as
\begin{equation}
D_{\rm dipole}(r) \simeq A m^2 \sqrt{\frac{\pi}{2m}} \frac{1}{r^{1/2}} e^{-mr},
\end{equation}
asymptotically for large $r$.

\subsubsection{Summary of three fit-functions}

Here, we summarize the three fit-functions in Table~\ref{tab:threeFunctions}.
In the coordinate space, the functional forms look rather different, {\it i.e.}, 
$K_1(mr)/r$, $e^{-mr}/r$, and $K_0(mr)$.
However, there are systematic relations 
in their asymptotic form and their momentum representation, 
as shown in Table~\ref{tab:threeFunctions}. 
In the coordinate space, the difference of the asymptotic form 
is just the power of the prefactor as $r^{-n/2}$ ($n=1, 2, 3$).

\begin{table}[h]
\begin{center}
\caption{\label{tab:threeFunctions}
Summary of the functional form candidates, 
$D_{\rm mass}(r), D_{\rm Yukawa}(r)$, and $D_{\rm dipole}(r)$, 
together with their asymptotic form and their momentum representation.
}
\begin{tabular}{llll}
\hline\hline
& Functional form & Asymptotic form &  Momentum space \\
\hline
$D_{\rm mass}$ & $ r^{-1}K_1(mr)$ & $r^{-3/2}e^{-mr}$ & $(p^2+m^2)^{-1}$ \\
$D_{\rm Yukawa}$ & $r^{-1} e^{-mr}$ & $r^{-1}e^{-mr}$ & $(p^2+m^2)^{-3/2}$ \\
$D_{\rm dipole}$ & $K_0(mr)$ & $r^{-1/2}e^{-mr}$ & $(p^2+m^2)^{-2}$ \\
\hline\hline
\end{tabular}
\end{center}
\end{table}

From the aspect of the space-time dimension, 
the Yukawa-type and dipole-type functions may be regarded as 
``low-dimensional" propagator forms.
As was already mentioned, the Yukawa-type function has a three-dimensional character, 
since the Yukawa function is obtained by the Fourier transformation of 
the massive propagator $(p^2+m^2)^{-1}$ in the three-dimensional space-time.
From this viewpoint, the dipole-type function as $K_0(mr)$ has a two-dimensional character, 
since this function is obtained by the Fourier transformation of 
$(p^2+m^2)^{-1}$ in the two-dimensional space-time as 
\begin{equation}
\int \frac{d^2p}{(2\pi)^2} e^{ip\cdot x}\frac{1}{p^2+m^2} = \frac{1}{2\pi}
K_0(mr).
\end{equation}
The derivation of this formula is shown in Appendix~A.

\subsection{Comparison of lattice QCD results with fit-functions}

In this subsection, we compare the lattice gluon propagator  
with the three fit-functions, 
$D_{\rm mass}(r)$, $D_{\rm Yukawa}(r)$, and $D_{\rm dipole}(r)$.
In the analysis, we consider the scalar-type gluon propagator $D(r)$ 
in the Landau gauge in the range of $r = 0.1 \sim 1.0$fm obtained in SU(3) lattice QCD,
and try to reproduce the lattice data 
through the fit-analysis with various range of $r$ for each fit-function.

\begin{figure}[h]
\begin{center}
\includegraphics[scale=1.1]{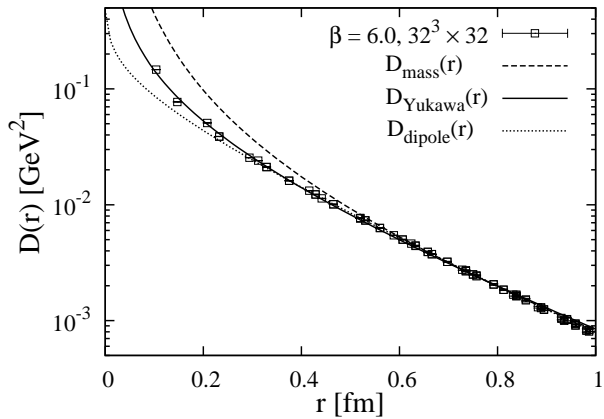}
\caption{\label{fig:threeFitting}
The fit result of the lattice gluon propagator $D(r)$ with the three functional forms,   
$D_{\rm Yukawa}(r)$ (solid line), 
$D_{\rm dipole}(r)$ (dotted line), and $D_{\rm mass}(r)$ (dashed line).
The Yukawa-type function $D_{\rm Yukawa}(r)$ 
well reproduces the lattice result at $\beta=6.0$ in the whole region of $r=0.1 \sim 1.0$fm.}
\end{center}
\end{figure}

Figure~\ref{fig:threeFitting} shows the typical example of 
the fit result of the lattice gluon propagator $D(r)$ at $\beta = 6.0$.
For each fit-function, the best-fit parameters ($m$, $A$) and the fit range 
are listed in Table~\ref{tab:threeFitting}.

\begin{table}[h]
\begin{center}
\caption{\label{tab:threeFitting}
The best-fit parameters ($m$, $A$) and the fit range 
in the fit analysis of the lattice gluon propagator $D(r)$ 
at $\beta = 6.0$ with the three functions, 
$D_{\rm mass}(r) = AmK_1(mr)/r$, $D_{\rm Yukawa}(r) = A m e^{-mr}/r$, and
$D_{\rm dipole}(r) = Am^2 K_0(mr)$.
}
\begin{tabular}{llll}
\hline\hline
Functional form ~& ~Fit range [fm]~~ & ~$m$ [GeV]~~~~ & ~~~~$A$~~~~ \\
\hline
$D_{\rm mass}$ & $0.6 \sim 1.0$ & 0.517(5) & 0.125(2)  \\
$D_{\rm Yukawa}$ &$0.1\sim 1.0$ & 0.624(8) & 0.162(2)  \\
$D_{\rm dipole}$ &$0.4\sim 1.0$ & 0.817(1) & 0.123(1)  \\
\hline\hline
\end{tabular}
\end{center}
\end{table}

As a remarkable fact, the Yukawa-type function $D_{\rm Yukawa}(r)$ 
well reproduces the lattice QCD data in the whole region of 
$r = 0.1 \sim 1.0$fm.
On the other hand, the dipole-type function $D_{\rm dipole}(r)$ 
fails to reproduce the whole region of the lattice data, 
since it gives too strong reduction at the short distance.
The massive propagator $D_{\rm mass}(r)$ also fails to 
reproduce the whole region of the lattice data, as was already shown. 

\begin{figure}[h]
\begin{center}
\includegraphics[scale=1.1]{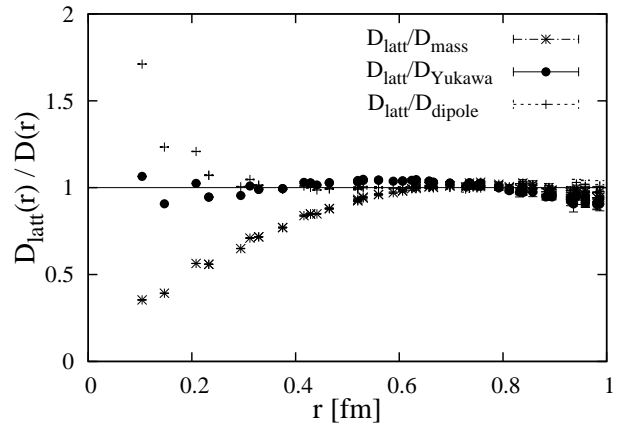}
\caption{\label{fig:threeFittingRatio}
The ratio of the lattice QCD data $D_{\rm latt}(r)$ at $\beta$ = 6.0  
to the fit-functions $D_{\rm mass}(r)$, $D_{\rm Yukawa}(r)$ and $D_{\rm dipole}(r)$ 
on the scalar-type gluon propagator, {\it i.e.}, 
$D_{\rm latt}/D_{\rm mass}$, $D_{\rm latt}/D_{\rm Yukawa}$, 
and $D_{\rm latt}/D_{\rm dipole}$.
}
\end{center}
\end{figure}

To see the difference of the three fit results clearer, 
we show in Fig.\ref{fig:threeFittingRatio} 
the ratio of the lattice QCD data $D_{\rm latt}(r)$ to the three fit-functions 
on the scalar-type gluon propagator, 
{\it i.e.}, $D_{\rm latt}/D_{\rm mass}$, $D_{\rm latt}/D_{\rm Yukawa}$, 
and $D_{\rm latt}/D_{\rm dipole}$. 
One finds $D_{\rm latt}/D_{\rm Yukawa} \simeq 1$ 
in the whole region of $r = 0.1 \sim 1$fm, 
while $D_{\rm latt}/D_{\rm mass}$ and $D_{\rm latt}/D_{\rm dipole}$
differ from the unity for small $r$.

\subsection{Yukawa-type gluon propagator in Landau gauge}

We thus find an appropriate functional form of the gluon propagator in the Landau gauge. 
In fact, the scalar-type gluon propagator $D(r)$ in the coordinate space 
is well described by the Yukawa-type function $D_{\rm Yukawa}(r)=Ame^{-mr}/r$ 
with $m=0.624(8)$GeV and $A=0.162(2)$ in the whole range of $r=0.1 \sim 1.0$fm.

\begin{figure}[h]
\begin{center}
\includegraphics[scale=1.1]{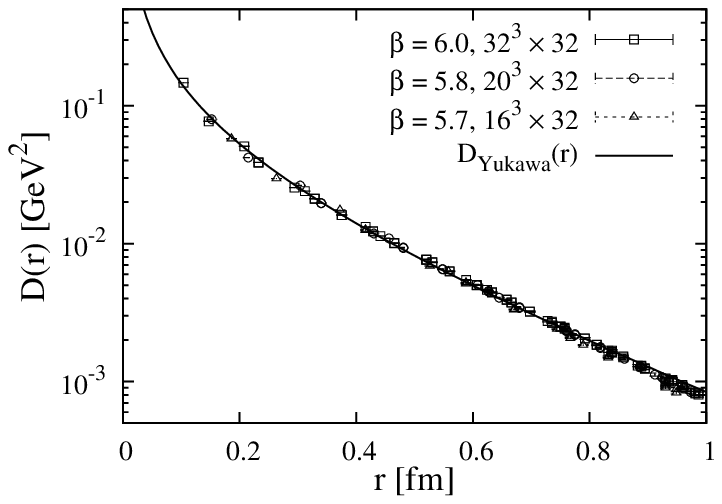}
\includegraphics[scale=1.1]{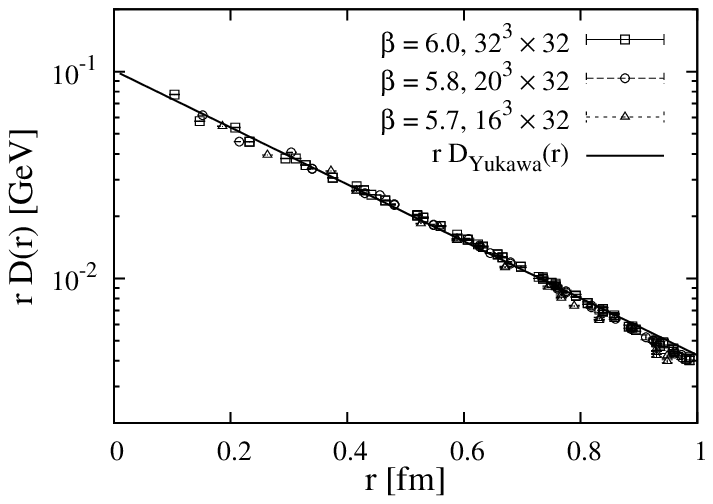}
\caption{\label{fig:YukawaSummary}
The Yukawa-type function 
$D_{\rm Yukawa}(r)=Ame^{-mr}/r$ (solid line) 
with $m=0.624$GeV and $A=0.162$
obtained by the fit analysis at $\beta=6.0$, 
and the lattice QCD data of the scalar-type gluon propagator $D(r)$ in the Landau gauge 
at $\beta$=5.7, 5.8, and 6.0 in the range of $r=0.1 \sim 1.0$fm.
The lower figure is the logarithmic plot of $rD(r)$ and $rD_{\rm Yukawa}(r)$. 
}
\end{center}
\end{figure}

As a summary figure, we show in Fig.\ref{fig:YukawaSummary} 
the comparison between the obtained Yukawa-type function $D_{\rm Yukawa}(r)$ 
and all the lattice QCD data of the scalar-type gluon propagator $D(r)$ 
at $\beta$=5.7, 5.8, and 6.0 in the range of $r=0.1 \sim 1.0$fm.
For clearer presentation of the Yukawa-functional behavior of the lattice gluon propagator $D(r)$, 
we also show the logarithmic plot of $rD(r)$ in Fig.\ref{fig:YukawaSummary}(b).
All the lattice QCD data at $\beta$=5.7, 5.8, and 6.0 are found to be 
well reproduced with the Yukawa-type function in the range of $r=0.1 \sim 1.0$fm.
Note also that all the lattice data of $rD(r)$ locate around a straight line  
in the logarithmic plot of Fig.\ref{fig:YukawaSummary}(b).

\begin{figure}[h]
\begin{center}
\includegraphics[scale=1.1]{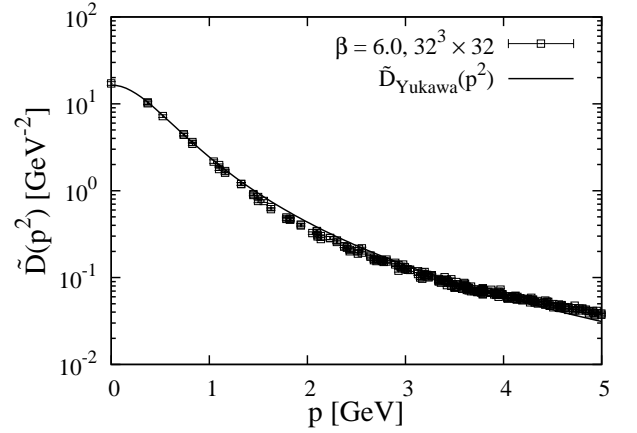}
\caption{\label{fig:YukawaMom}
The Yukawa-type propagator in the momentum space, {\it i.e.}, 
$\tilde{D}_{\rm Yukawa}(p^2) = 4\pi^2 Am (p^2+m^2)^{-3/2}$ (solid line) 
with $m = 0.624$GeV and $A= 0.162$, 
the same values used in Fig.\ref{fig:YukawaSummary}. 
The horizontal axis is $p \equiv (p_\alpha p_\alpha)^{1/2}$.
The symbols denote the lattice QCD data of 
the scalar-type gluon propagator $\tilde{D}(p^2)$ in the Landau gauge  
at $\beta =6.0$, where the momentum is defined as $p_\mu=\frac{2}{a}\sin(\frac{\pi n_\mu}{L_\mu})$.
}
\end{center}
\end{figure}

Next, we investigate the gluon propagator $\tilde D(p^2)$ in the momentum space 
in terms of the Yukawa-type function.
In Fig.\ref{fig:YukawaMom}, we show the scalar-type gluon propagator $\tilde D(p^2)$ 
in the Landau gauge obtained in lattice QCD at $\beta = 6.0$, 
and $\tilde{D}_{\rm Yukawa}(p^2) = 4\pi^2 Am (p^2+m^2)^{-3/2}$, which is 
the Fourier transformation of the Yukawa-type function $D_{\rm Yukawa}(r)$. 
The horizontal axis is $p \equiv (p_\alpha p_\alpha)^{1/2}$.
Here, we use the same parameters $m = 0.624$GeV and $A= 0.162$ as 
those used in the best-fit analysis for the coordinate-space gluon propagator.
From Figs.~\ref{fig:latResMom} and \ref{fig:YukawaMom},
the lattice QCD data of $\tilde{D}(p^2)$ at $\beta$=5.7, 5.8, and 6.0 
are found to be approximated with $\tilde{D}_{\rm Yukawa}(p^2)$ 
in the range of $p \le 3{\rm GeV}$.
We also perform the best-fit analysis of the lattice data of 
the gluon propagator $\tilde{D}(p^2)$ in the momentum space 
with $\tilde{D}_{\rm Yukawa}(p^2) = 4\pi^2 Am (p^2+m^2)^{-3/2}$.
For the lattice data at $\beta$=6.0 in the fit range of $p^2 \le (3{\rm GeV})^2$, 
the best-fit parameters are found to be $m \simeq 0.577$GeV and $A \simeq 0.151$, 
which are close to the values obtained from 
the fit-analysis of the coordinate-space gluon propagator.

As a caution, in the UV region of $p > 3{\rm GeV}$, 
the gluon propagator $\tilde D(p^2)$ gradually deviates 
from the Yukawa form $D_{\rm Yukawa}(p^2)$ 
and gradually approaches the perturbative form $1/p^2$.
However, the coordinate-space gluon propagator $D(r)$ is found to be 
almost unchanged for $r=0.1 \sim 1.0$fm by correcting the UV deviation.

Also in the Deep-IR region of $p < 0.5{\rm GeV}$, 
as is briefly summarized in Appendix~B, 
there appears some deviation 
between $\tilde D(p^2)$ and $\tilde D_{\rm Yukawa}(p^2)$. 
In fact, such a Deep-IR deviation is indicated by 
recent huge-volume lattice QCD studies \cite{BIMPS0907,SO0607,CM0708}.
In the momentum space, the true gluon propagator 
$\tilde D(p^2)$ turns out to take a saturated value 
smaller than the Yukawa-type propagator 
$\tilde D_{\rm Yukawa}(p^2)$ 
in the Deep-IR region of $p < 0.5$GeV.
In other words, $p \simeq 0.5$GeV is the lower bound on 
the applicability of the Yukawa-type propagator $\tilde D_{\rm Yukawa}(p^2)$ 
to the gluon propagator.
Based on the huge-volume lattice data \cite{BIMPS0907}, 
we investigate the Deep-IR-corrected gluon propagator 
in the coordinate space in Appendix~B.
As the conclusion, even after the correction in the Deep-IR region, 
the Yukawa-type function is found to work well 
for $r = 0.1 \sim 1.0$fm. (See Fig.\ref{fig:hugeLatticeCorr} in Appendix~B.)

As the main conclusion of this paper, 
we summarize the functional form of the gluon propagator in the Landau gauge 
obtained in SU(3) lattice QCD.
\begin{enumerate}
\item
The coordinate-space gluon propagator $D(r)$ in the Landau gauge is 
well described by the four-dimensional Yukawa-type function as 
\begin{equation}
D(r) \equiv \frac{1}{24} D_{\mu\mu}^{aa}(r)= A \frac{m}{r} e^{-mr},
\end{equation}
with $m \simeq$ 600MeV and $A \simeq$ 0.16,
for the whole region of $r \equiv (x_\alpha x_\alpha)^{1/2}=0.1\sim 1.0$fm.
(This is valid even after the possible correction in the UV and Deep-IR regions.)
\item
The gluon propagator $\tilde{D}(p^2)$ in the momentum space 
is also well described by the corresponding new-type propagator 
(four-dimensional Fourier transformed Yukawa-type function) as 
\begin{eqnarray}
\tilde{D}(p^2) =\frac{1}{24} \tilde{D}_{\mu\mu}^{aa}(p^2)= \frac{4\pi^2Am}{(p^2+m^2)^{3/2}},
\end{eqnarray}
with $m \simeq$ 600MeV and $A \simeq$ 0.16 (same values),
in the momentum region of $0.5{\rm GeV} \le p \le 3{\rm GeV}$.
\end{enumerate}
Note here that all the component of the gluon propagator 
$D_{\mu\nu}^{ab}(x-y)=\langle A_\mu^a(x) A_\nu^b(y) \rangle$ 
and $\tilde{D}_{\mu\nu}^{ab}(p)$ in the Landau gauge 
can be analytically expressed, starting from the Yukawa-type function. (See Sec.II.)

Quantitatively, the Yukawa-type propagator $D(r)$ exhibits a slower decreasing feature, 
and $\tilde D(p^2)$ exhibits faster decreasing, 
in comparison with the ordinary massive propagator.
It seems suggestive to rewrite $\tilde D(p^2)$ as  
\begin{equation} 
\tilde D(p^2)=\frac{Z(p^2)}{p^2+m^2}, \quad
Z(p^2)= \frac{4\pi^2Am}{(p^2+m^2)^{1/2}},
\end{equation}
where $Z(p^2)$ corresponds to the wave-function renormalization of the gluon field, 
in a similar manner to the Schwinger-Dyson formalism. 
Near the on-shell-like condition of $p^2+m^2=0$, $Z(p^2)$ tends to diverge as $+\infty$, 
which leads to anomalous gluon propagation and 
may mimic the gluon confinement, {\it i.e.},  the absence of on-shell gluon states. 

To see the physical meaning of the massive parameter $m$ in the Yukawa-type function,  
we compare the Yukawa-type propagator form $e^{-mr}/r$ 
with the massive propagator form $K_1(mr)/r$ in the coordinate space.
In spite of a significant difference at the short distance, 
their difference is just the prefactor at the large distance, 
where $K_1(mr)/r \sim e^{-mr}/r^{3/2}$ and the main reduction factor is $e^{-mr}$.
Then, the mass parameter $m \simeq$ 600MeV in the Yukawa-type gluon propagator 
directly corresponds to the effective gluons mass 
in the infrared region. (See Sec.VI C.)

Here, we briefly comment on the other functional forms for the gluon propagator. 
Up to now, many functional forms of the gluon propagator $\tilde D(p^2)$ 
have been considered and compared with the lattice QCD result 
\cite{C8207,G78,Z91920204,S868790,MMST9395,UK9899,CZ02,IOS05}.
Some functional forms well describe the gluon propagator better than the Yukawa-type function,
but they need four or more fit parameters and take highly non-analytical complicated form. 
On the other hand, the Yukawa-type function $D_{\rm Yukawa}(r)$ 
has {\it only two parameters} $(A,m)$ and takes an analytical form, which are its advantages.
For example, owing to the analyticity, the Yukawa-type gluon propagator $D_{\rm Yukawa}(r)$ 
leads to analytical applications of gluonic nonperturbative quantities, 
as will be demonstrated in the next section.

\section{\label{sec:effmassanalytic} 
Analytical applications of Yukawa-type gluon propagator}

In this section, as the applications of the Yukawa-type gluon propagator, 
we derive analytical expressions for the zero-spatial-momentum propagator $D_0(t)$, 
the effective mass $M_{\rm eff}(t)$, and the spectral function $\rho(\omega)$ of the gluon field. 
All the derivations can be analytically performed, 
starting from the Yukawa-type gluon propagator $D_{\rm Yukawa}(r)$.

\subsection{Zero-spatial-momentum propagator of gluons}

Second, we consider the zero-spatial-momentum propagator $D_0(t)$, 
associated with the Yukawa-type propagator $D_{\rm Yukawa}(r)$.
$D_0(t)$ was introduced in Sec.IV-B 
in the context of the effective mass in lattice QCD. 
Here, we mainly deal with the continuum formalism with infinite spatial volume.
For the simple argument, we first neglect the temporal periodicity, 
which is justified for large temporal lattice size. 

We start from the Yukawa-type gluon propagator, 
\begin{equation}
D_{\rm Yukawa}(r) =\frac{Am}{r} e^{-mr}
=\frac{Am}{\sqrt{\vec{x}^2+t^2}} e^{-m\sqrt{\vec{x}^2+t^2}},
\end{equation}
with $r = \sqrt{\vec{x}^2+t^2}$. 
Like Eq.(\ref{eq:ZMPcorrL}),
the zero-spatial-momentum propagator is given by 
\begin{equation}
D_0(t) \equiv \frac{1}{24} \int d^3 x 
\langle A_\mu^a(\vec{x},t)A_\mu^a(\vec{0},0) \rangle
=\int d^3 x \ D_{\rm Yukawa}(r)
\label{eq:contZMP}
\end{equation}
in the continuum formalism. 
Using the three-dimensional polar coordinate of $\vec x$, 
we calculate this integral as follows:
\begin{eqnarray}
D_0(t) &=& 4\pi Am \int_0^\infty dx \ x^2 \frac{1}{\sqrt{x^2+t^2}}e^{-m\sqrt{x^2+t^2}} 
\nonumber \\
&=& 4 \pi A m \int_t^\infty dr \sqrt{r^2-t^2} e^{-mr} 
\nonumber \\
&=& 4\pi A m t^2 \int_1^\infty d\bar{r} \sqrt{\bar{r}^2-1}e^{-\bar{r}mt} 
\nonumber \\ 
&=& 4\pi A m t^2 \frac{1}{mt} K_1(mt) = 4\pi At K_1(mt),
\end{eqnarray}
with $\bar{r} \equiv r/t$. 
Here, we have used Eq.(\ref{eq:IntegFormModBessel}) 
on the modified Bessel function. 
Thus, we derive an analytical expression for  
the zero-spatial-momentum propagator, 
\begin{equation}
D_0(t) = 4\pi A t K_1(mt).
\label{eq:D0Yukawa}
\end{equation}

\begin{figure}[h]
\begin{center}
\includegraphics[scale=1.1]{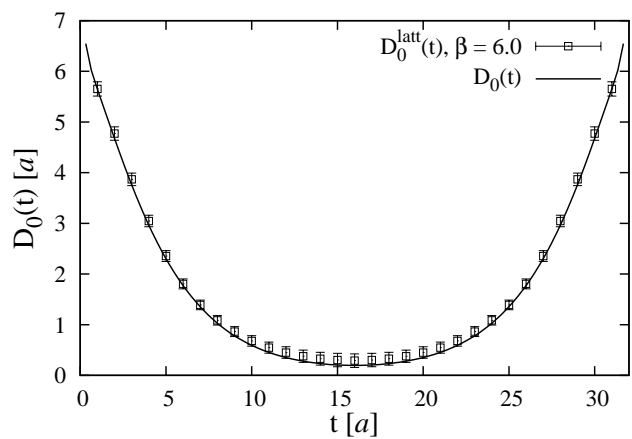}
\caption{\label{fig:CorrYukawa}
The zero-spatial-momentum propagator $D_0(t)$ of gluons in the Landau gauge.
The symbols are the lattice QCD data at $\beta = 6.0$,
and the solid line denotes the theoretical curve of 
Eq.(\ref{eq:corrAnalytical}), derived 
from the Yukawa-type propagator with $m$=0.624GeV and $A$=0.162, 
the same values in Fig.\ref{fig:YukawaSummary}.
}
\end{center}
\end{figure}

For the actual comparison with the lattice QCD data, 
we take account of the temporal periodicity, 
which is used in lattice calculations.
In this case, $D_0(t)$ is given as
\begin{equation}
D_0(t) = 4\pi A [ t K_1(mt) + \left(N_t - t\right) K_1(m(N_t-t))].
\label{eq:corrAnalytical}
\end{equation}
In Fig.\ref{fig:CorrYukawa}, we show the lattice QCD result of $D_0(t)$ in the Landau gauge, 
and the theoretical curve of Eq.(\ref{eq:corrAnalytical}) 
with $m$=0.624GeV and $A$=0.162, the same values used in the previous section.
The lattice QCD data of $D_0(t)$ are found to be well described by  
the theoretical curve, associated with the Yukawa-type gluon propagator. 

We also consider $D_0(t)$ in the lattice formalism, 
for more direct comparison with the lattice QCD data.
Here, we use the expression of $D_0(t)$ with $\tilde D_{\rm Yukawa}(p^2)$,
\begin{eqnarray}
D_0(t) &=& \int d^3 x \int \frac{d^4 p}{(2\pi)^4} e^{ip\cdot x} \tilde{D}_{\rm Yukawa}(p^2) 
\nonumber \\
&=& \int_{-\infty}^{\infty} \frac{dp_0}{2\pi} e^{ip_0 t} \tilde{D}_{\rm Yukawa}(p_0^2).
\end{eqnarray}
On the lattice, $\tilde{D}_{\rm Yukawa}(p_0^2)\propto (p_0^2+m^2)^{-3/2}$ is given as
\begin{eqnarray}
\tilde{D}_{\rm Yukawa}(p_0^2)= 
\frac{4\pi^2 A m}{\big\{ \big( 2\sin \big( \frac{\pi n}{N_t} \big) \big)^2 + m^2 \big\}^{3/2}},
\end{eqnarray}
with $p_0=2\sin(\pi n/N_t)$ ($n=0,1,2, ... , N_t -1$) in the lattice unit.
Then, we obtain an analytical expression for 
the zero-spatial-momentum propagator,
\begin{equation}
D_0(t) = \frac{1}{N_t}\sum_{n=0}^{N_t-1} e^{i \frac{2\pi n}{N_t}t}
\frac{4\pi^2 A m}{\big\{ \big( 2\sin \big( \frac{\pi n}{N_t} \big) \big)^2 + m^2 \big\}^{3/2}}.
\end{equation}
No significant numerical difference turns out to 
be found between the continuum and the lattice formulae, 
under the condition of our lattice QCD calculation.
This means that the integral quantity $D_0(t)$ is not so sensitive to the details of 
the UV-behavior of the propagator $D_{\rm Yukawa}(r)$.

In the UV region, the Yukawa-type propagator $\tilde D_{\rm Yukawa}(p^2) \propto (p^2+m^2)^{-3/2}$ 
deviates from the correct behavior of the perturbative propagator $1/p^2$.
For the quantitative estimate of the influence from the deviation in the UV region, 
we calculate the zero-spatial-momentum propagator $D_0(t)$ 
using the UV-corrected Yukawa propagator $\tilde D_{\rm Yukawa}^{\rm UVcorr}(p^2)$, 
which is $\tilde D_{\rm Yukawa}(p^2)$ for $p \le 4$GeV and $1/p^2$ in the UV region of $p \ge 4$GeV.
The difference of $D_0(t)$ is found to be at most 1\% 
between the cases with $\tilde D_{\rm Yukawa}(p^2)$ and $\tilde D_{\rm Yukawa}^{\rm UVcorr}(p^2)$.
Thus, the integral quantity $D_0(t)$ is insensitive to the UV behavior of the gluon propagator, 
so that it is also expected to be insensitive to the discritization error \cite{UK9899,B9900,dSR07}
in the UV region in the lattice calculation.

\subsection{Effective mass of gluons}

Second, we investigate the effective mass $M_{\rm eff}(t)$, 
as the consequence of the Yukawa-type propagator $D_{\rm Yukawa}(r)$.
The effective-mass plot is a  general useful technique for the mass estimation in lattice QCD, 
and was actually examined for gluons in Sec.IV-B.
For simplicity, we here treat the three-dimensional space as a continuous infinite-volume space, 
while the temporal variable $t$ is discrete and periodic.
For the simple notation, we here use the lattice unit for $t$. 

When the temporal periodicity can be neglected, 
the zero-spatial-momentum propagator $D_0(t)$ is expressed by Eq.(\ref{eq:D0Yukawa}), 
and we obtain an analytical expression of the effective mass,  
\begin{equation}
M_{\rm eff}(t) = \ln \frac{D_0(t)}{D_0(t+1)}
=\ln \frac{t K_1(mt)}{(t+1)K_1(m(t+1))}.
\label{eq:EMGYukawa}
\end{equation}
From the asymptotic form of $K_1(z) \propto z^{-1/2}e^{-z}$, 
the effective mass of gluons is approximated as
\begin{equation}
M_{\rm eff}(t) \simeq 
m - \frac{1}{2} \ln \big( 1 + \frac{1}{t} \big)
\simeq  m - \frac{1}{2t}
\label{eq:meffAsymptotic}
\end{equation}
for large $t$. This functional form indicates that 
$M_{\rm eff}(t)$ is an increasing function and approaches $m$ 
from below, as $t$ increases.

Note that the mass parameter $m$ in the Yukawa-type gluon propagator 
directly corresponds to the effective mass $M_{\rm eff}(t)$ 
of gluons for large $t$. In fact, $m \simeq$ 600MeV has 
a definite physical meaning of the effective gluon mass in the infrared region.

Note also that the simple analytical expression of Eq.(\ref{eq:EMGYukawa}) or (\ref{eq:meffAsymptotic}) 
reproduces the anomalous increasing behavior of the effective mass $M_{\rm eff}(t)$ of gluons, 
as was observed in Fig.\ref{fig:effMass}. 
As for the increasing behavior, there is a general argument: 
this can occur when the spectral function is not positive definite 
\cite{MO87,BPS94,MMST9395,M99}, 
although the concrete form of the gluon spectral function is not yet known.
Instead, this framework with the Yukawa-type gluon propagator  
gives an analytical and quantitative method, and 
is found to well reproduce the lattice result.
(The actual comparison is demonstrated with Eq.(\ref{eq:meffAnaytical}).)

Next, we take account of the temporal periodicity, which is used in lattice QCD calculations.
In this case, the effective mass $M_{\rm eff}(t)$ is defined by ``cosh-type" as Eq.(\ref{eq:EMcosh}), 
and the zero-spatial-momentum propagator $D_0(t)$ is given by Eq.(\ref{eq:corrAnalytical}).
Then, the effective mass $M_{\rm eff}(t)$ of gluons is expressed as 
\begin{eqnarray}
\label{eq:meffAnaytical}
&& \frac{\cosh\left[ M_{\rm eff}(t) (N_t/2 - (t+1))\right]}
{\cosh\left[ M_{\rm eff}(t) (N_t/2-t)\right]}=
\\
&&\frac{(t+1)K_1(m(t+1)) + (N_t -(t+1))K_1(m(N_t-(t+1)))}{tK_1(mt)+(N_t-t)K_1(m(N_t-t))}.
\nonumber
\end{eqnarray}

\begin{figure}[h]
\begin{center}
\includegraphics[scale=1.1]{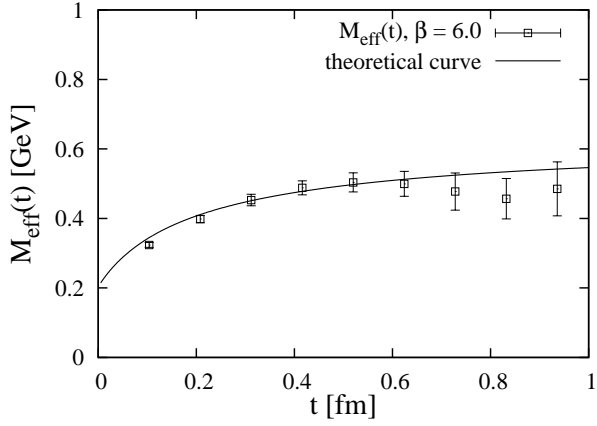}
\caption{\label{fig:effMassYukawa}
The effective mass $M_{\rm eff}(t)$ of gluons in the Landau gauge.
The symbols denote the lattice QCD data at $\beta = 6.0$, 
and the solid line denotes the theoretical curve of 
Eq.(\ref{eq:meffAnaytical}) 
derived from the Yukawa-type propagator with $m$=0.624GeV,
the same value used in Fig.\ref{fig:YukawaSummary}.
}
\end{center}
\end{figure}

In Fig.\ref{fig:effMassYukawa}, we show the theoretical curve obtained by Eq.(\ref{eq:meffAnaytical}) 
together with the lattice result of $M_{\rm eff}(t)$.
Here, we take $m$=0.624GeV, the same value used in the previous section.
The lattice QCD data of $M_{\rm eff}(t)$ are found to be well described by  
the theoretical curve, associated with the Yukawa-type gluon propagator. 

\subsection{Spectral function of gluons in Landau gauge}

From the analytical expression of the zero-spatial-momentum propagator  
$D_0(t)=4\pi A t K_1(mt)$ in Eq.(\ref{eq:D0Yukawa}), 
we can derive the spectral function $\rho(\omega)$ of the gluon field, 
associated with the Yukawa-type gluon propagator.
For simplicity,
we take continuum formalism with infinite space-time.

The relation between the spectral function $\rho(\omega)$ and 
the zero-spatial-momentum propagator $D_0(t)$ 
is given by the Laplace transformation, 
\begin{equation}
D_0(t) = \int_0^\infty d\omega \ \rho(\omega) \ e^{-\omega t}.
\label{eq:CorrvsSF}
\end{equation}
When the spectral function is given by a $\delta$-function 
such as $\rho(\omega) \sim \delta(\omega-\omega_0)$, 
which corresponds to a single mass spectrum, 
one finds a familiar relation of $D_0(t) \sim e^{-\omega_0t}$.
For the physical state, the spectral function $\rho(\omega)$ gives 
a probability factor, and is non-negative definite in the whole region of $\omega$.
This property is related to the unitarity of the $S$-matrix.

In general, the Laplace transformation is defined by 
\begin{equation}
g(t) = \int_0^\infty d\omega \ e^{-\omega t} \ f(\omega),
\end{equation}
and the inverse Laplace transformation is expressed as 
\begin{equation}
f(\omega) = \frac{1}{2\pi i}\int_{c-i\infty}^{c+i\infty} dt \ e^{\omega t} \ g(t).
\end{equation}

Then, from Eq.(\ref{eq:CorrvsSF}), the spectral function $\rho(\omega)$ is expressed as 
\begin{eqnarray}
\rho(\omega)&=&\frac{1}{2\pi i}\int_{c-i\infty}^{c+i\infty} dt \ e^{\omega t} \ D_0(t)
\nonumber \\
&=&\frac{1}{2\pi i}\int_{c-i\infty}^{c+i\infty} dt \ e^{\omega t} \ 4\pi A t K_1(mt)
\nonumber \\
&=&\frac{1}{2\pi i}\int_{c'-i\infty}^{c'+i\infty} dt' \ e^{{\omega}' t'} \ \frac{4\pi A}{m^2} t' K_1(t'),
\label{eq:SFint}
\end{eqnarray}
with ${\omega}' \equiv \omega/m$, $t' = mt$ and $c' = mc$.

Performing the partial integration in 
Eq.(\ref{eq:IntegFormModBessel}) on the modified Bessel function, 
we obtain a formula of the Laplace transformation,
\begin{eqnarray}
K_1(t) &=& \int_1^\infty d\omega \ e^{-\omega t} \frac{\omega}{(\omega^2-1)^{1/2}}
\nonumber \\
&=& \int_0^\infty d\omega \ e^{-\omega t} \frac{\omega}{(\omega^2-1)^{1/2}}\theta(\omega-1),
\end{eqnarray}
which leads to the inverse Laplace transformation,
\begin{equation}
\frac{1}{2\pi i}\int_{c-i\infty}^{c+i\infty} dt \ e^{\omega t} \ K_1(t)
=\frac{\omega}{(\omega^2-1)^{1/2}} \theta(\omega-1).
\label{eq:LTK1}
\end{equation}
By differentiating this formula by $\omega$, we find  
\begin{eqnarray}
& &\frac{1}{2\pi i}\int_{c-i\infty}^{c+i\infty} dt \ e^{\omega t} \ t K_1(t)
\nonumber \\
&=&-\frac{1}{(\omega^2-1)^{3/2}} \theta(\omega-1) 
+\frac{\omega}{(\omega^2-1)^{1/2}} \delta(\omega-1) 
\nonumber \\
&=&-\frac{1}{(\omega^2-1)^{3/2}} \theta(\omega-1) 
+\frac{1}{\{2(\omega-1)\}^{1/2}} \delta(\omega-1),~~~~~~~
\end{eqnarray}
where the second term includes an infinite factor 
besides the $\delta$-function.
Then, we apply this formula to Eq.(\ref{eq:SFint}), 
and obtain the spectral function $\rho(\omega)$ as 
\begin{eqnarray}
& & \rho(\omega)=\frac{1}{2\pi i}\int_{c'-i\infty}^{c'+i\infty} 
dt' \ e^{\omega' t'} \ \frac{4\pi A}{m^2} t' K_1(t'),
\nonumber \\
&=&-\frac{4\pi A/m^2}{({\omega'}^2-1)^{3/2}} \theta(\omega'-1) 
+\frac{4\pi A/m^2}{\{2(\omega'-1)\}^{1/2}} \delta(\omega'-1)
\nonumber \\
&=&-\frac{4\pi A m}{(\omega^2-m^2)^{3/2}} \theta(\omega-m) 
+\frac{4\pi A/\sqrt{2m}}{(\omega-m)^{1/2}} \delta(\omega-m).~~~~~~~
\end{eqnarray}

For more rigorous derivation, we avoid the singularity at $\omega=m$ 
by regularizing Eq.(\ref{eq:LTK1}) as  
\begin{equation}
\frac{1}{2\pi i}\int_{c-i\infty}^{c+i\infty} dt \ e^{\omega t} \ K_1(t)
=\frac{\omega}{(\omega^2-1)^{1/2}} \theta(\omega-1-\varepsilon) 
\end{equation}
with a positive infinitesimal $\varepsilon$, and find the formula of 
\begin{eqnarray}
&& \frac{1}{2\pi i}\int_{c-i\infty}^{c+i\infty} dt \ e^{\omega t} \ t K_1(t)
\nonumber \\
&&=-\frac{1}{(\omega^2-1)^{3/2}} \theta(\omega-1-\varepsilon) 
+\frac{1}{(2\varepsilon)^{1/2}} \delta(\omega-1-\varepsilon),~~~~~~~
\end{eqnarray}
which leads to the regularized spectral function, 
\begin{equation}
\rho_\varepsilon(\omega)
=-\frac{4\pi A m}{(\omega^2-m^2)^{3/2}}\theta(\omega-m-\varepsilon)
+\frac{4\pi A}{(2m\varepsilon)^{1/2}} \delta(\omega-m-\varepsilon).
\label{eq:SFYukawaR}
\end{equation}
In the calculation of the Laplace transformation from $\rho(\omega)$ to $D_0(t)$, 
we can avoid the divergence of the integral at $\omega=m$ 
by using the regularized spectral function $\rho_\varepsilon(\omega)$. 
Then, we can perform the integration, and properly obtain $D_0(t)=4\pi A tK_1(mt)$, 
by taking the limit of $\varepsilon \rightarrow 0$ after the integration. 
We have numerically confirmed that the spectral function $\rho_{\varepsilon}(\omega)$ 
in Eq.(\ref{eq:SFYukawaR}) reproduces $D_0(t)=4\pi A tK_1(mt)$ 
by the Laplace transformation in the limit of $\varepsilon \rightarrow 0$. 

In this way, we derive the spectral function $\rho(\omega)$ of the gluon field, 
associated with the Yukawa-type propagator:
\begin{equation}
\rho(\omega)
= -\frac{4\pi A m}{(\omega^2-m^2)^{3/2}}\theta(\omega-m)
+\frac{4\pi A/\sqrt{2m}}{(\omega-m)^{1/2}} \delta(\omega-m),
\label{eq:SFYukawa}
\end{equation}
which is regularized as $\rho_\varepsilon(\omega)$ in more rigorous derivation.  
Here, $m \simeq$ 600MeV is the mass parameter in the Yukawa-type function  
for the Landau-gauge gluon propagator.
The first term expresses a negative continuum spectrum, and 
the second term a $\delta$-functional peak with 
the residue including an infinite factor, which is positive as $\varepsilon^{-1/2}$ 
at $\omega=m+\varepsilon$.

\begin{figure}[h]
\begin{center}
\includegraphics[scale=1.1]{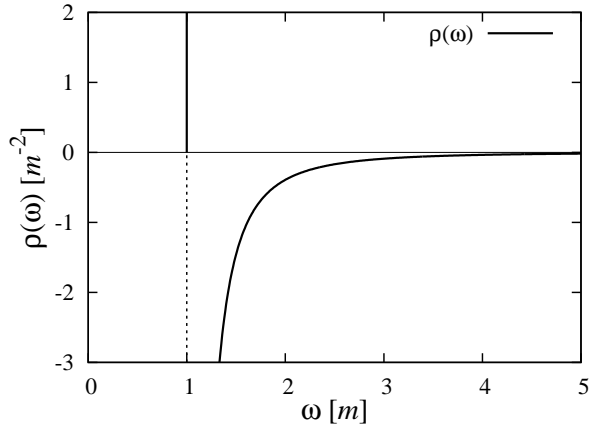}
\caption{\label{fig:Spectral}
The spectral function $\rho(\omega)$ of the gluon field, 
associated with the Yukawa-type propagator.
The unit is normalized by the mass parameter $m \simeq$ 600MeV.
As Eq.(\ref{eq:SFYukawa}) indicates, 
$\rho(\omega)$ shows anomalous behaviors: 
it has a positive $\delta$-functional peak with the residue of $+\infty$ at $\omega = m \ ( + \varepsilon )$, 
and takes negative values for all the region of $\omega > m$.
}
\end{center}
\end{figure}

We show in Fig.\ref{fig:Spectral} the spectral function $\rho(\omega)$ of the gluon field.
Although the appearance of the negative-value region in the gluon spectral function 
is expected, $\rho(\omega)$ exhibits two anomalous behaviors:
it has a positive $\delta$-functional peak with the residue of $+\infty$ at $\omega = m \ ( + \varepsilon )$, 
and it takes negative values for all the region of $\omega > m$.
This negative contribution of the spectral function gives 
unusual behavior of the effective mass $M_{\rm eff}(t)$ of gluons, 
{\it i.e.}, its increasing behavior on $t$.

As was discussed in Sec.IV, if the spectral function is non-negative definite, like the ordinary hadronic correlator, 
the effective mass must be a monotonously decreasing function, 
due to the mathematical nature of the summation of $e^{-\omega t}$ with non-negative coefficients, 
as can be proven with Eq.(\ref{eq:dMdt}). 
Physically, this is due to a larger positive contribution of the excited states 
to the effective mass at a shorter distance.
However, if the spectral function includes the negative-value region, 
such a definite tendency is lost \cite{M99,MO87,GGKPSW87,BPS94,MMST9395}.
This is the case of the gluon field in the Landau gauge.

More precisely, from Eq.(\ref{eq:dMdt}), we find that the increasing property of $M_{\rm eff}(t)$ can be realized, 
only when there is some suitable coexistence of positive- and negative-value regions in the spectral function $\rho(\omega)$.
As a remarkable fact, the obtained gluon spectral function $\rho(\omega)$ is 
negative-definite for all the region of $\omega > m$, except for the positive $\delta$-functional peak at $\omega=m$.
The negative property of the spectral function in coexistence with the positive peak 
leads to the anomalous increasing behavior of the effective mass $M_{\rm eff}(t)$.

In the UV region, the true gluon propagator deviates from the Yukawa-type function.
Then, we estimate the contribution from the UV part of this spectral function $\rho(\omega)$ 
to the zero-spatial-momentum propagator $D_0(t)$.
Introducing the UV cutoff $\Lambda$, we define the integration
\begin{equation}
D_0^{\Lambda}(t) = \int_0^\Lambda \rho(\omega) e^{-\omega t},
\end{equation}
which reduces to the Laplace transformation in the $\Lambda \rightarrow \infty$ limit.
For $t >$ 0.1fm, the $\Lambda$-dependence of the integral quantity $D_0^{\Lambda}(t)$ 
is found to be negligible for $\Lambda > 2$ GeV, {\it i.e.}, $D_0^{\Lambda}(t) \simeq D_0(t)$.
In other words, due to this UV-insensitivity, the obtained information on the UV part of 
the spectral function suffers from a large uncertainty, 
as is also seen in the maximum entropy method (MEM) analysis.
On the other hand, the IR/IM part of the spectral function $\rho(\omega)$ is relatively stable and reliable.

We note that the gluon spectral function $\rho(\omega)$ is divergent at $\omega=m+\varepsilon$, 
and the divergence structure is complicated and consists of two ingredients: 
a $\delta$-functional peak with a positive infinite residue and a negative wider power-damping peak.
On the finite-volume lattice, these singularities are to be smeared, 
and $\rho(\omega)$ is expected to take a finite value everywhere on $\omega$. 
On the lattice, we conjecture that the spectral function $\rho(\omega)$ includes  
a narrow positive peak stemming from the $\delta$-function in the vicinity of $\omega=m \ (+\varepsilon)$  
and a wider negative peak near $\omega \simeq m$ in the region of $\omega > m$.

In this way, the Yukawa-type gluon propagator suggests an extremely anomalous spectral function 
of the gluon field in the Landau gauge. 
Note that this framework gives an analytical and concrete expression 
for the gluon spectral function $\rho(\omega)$ at the quantitative level.
Actually, the resulting effective mass $M_{\rm eff}(t)$ 
well describes the lattice result, as shown in Fig.\ref{fig:effMass}.
The obtained gluon spectral function $\rho(\omega)$ is negative almost everywhere, and 
includes a complicated divergence structure near the ``anomalous threshold", $\omega=m\ (+\varepsilon)$. 

These anomalous features of the gluon spectral function 
may have some relation to the various nonperturbative QCD phenomena, 
such as the gluon confinement and the gluonic instability of the QCD vacuum, 
{\it e.g.}, gluon condensation, the Savvidy vacuum \cite{S77}, and the Copenhagen vacuum \cite{NO78AO80}.
In any case, the Yukawa-type gluon propagator $D_{\rm Yukawa}(r)$, which models 
the Landau-gauge gluon propagator, would be useful for 
the analytical and quantitative investigation of nonperturbative QCD.

\section{Summary and discussions}

We have studied the gluon propagator $D_{\mu\nu}^{ab}(x)$ 
in the Landau gauge in the infrared/intermediate region of $r \equiv (x_\mu x_\mu)^{1/2}=0.1 \sim$ 1.0fm, 
which is relevant to the quark-hadron physics,  
in SU(3) lattice QCD at $\beta$ = 5.7, 5.8, and 6.0 at the quenched level.
From the gluon propagator analysis and the effective-mass plot, 
we have estimated the effective gluon mass of $400 \sim 600$MeV 
in the infrared region of $r = 0.5 \sim 1.0$fm.
The effective gluon mass exhibits a significant $r$-dependence in this region: 
it takes a smaller value in the smaller $r$-region.

We have also studied the functional form of the gluon propagator $D_{\mu\nu}^{ab}(x)$ in lattice QCD. 
As a remarkable fact, the lattice QCD result 
of the Landau-gauge gluon propagator $D_{\mu\mu}^{aa}(r)$ is fairly well described 
by the Yukawa-type form $D_{\rm Yukawa}(r) \propto e^{-mr}/r$ with the mass parameter $m \simeq 600$MeV 
in the whole region of $r = 0.1 \sim 1.0$fm in four-dimensional Euclidean space-time.
This Yukawa-type propagator corresponds to the new-type propagator 
$\tilde D_{\rm Yukawa}(p^2) \propto (p^2+m^2)^{-3/2}$ in the momentum space, 
through the Fourier transformation, and this also well describes 
the lattice QCD result of the gluon propagator $\tilde D_{\mu\mu}(p^2)$ in the momentum space.

As the application of the Yukawa-type gluon propagator,
we have derived the analytical expression of 
the zero-spatial-momentum propagator $D_0(t)$, and the effective mass $M_{\rm eff}(t)$.
The obtained analytical functions for $D_0(t)$ and $M_{\rm eff}(t)$ 
well reproduce the lattice QCD results, in particular the anomalous increasing behavior of $M_{\rm eff}(t)$.
We have found that the mass parameter $m$ of the Yukawa-type gluon propagator 
directly corresponds to the effective gluon mass in the infrared region of $\sim$1fm.

We have also derived the analytical expression of the spectral function 
$\rho(\omega)$ of the gluon field, associated with the Yukawa-type gluon propagator, 
using the inverse Laplace transformation of the temporal propagator $D_0(t)$.
As a remarkable fact, the obtained spectral function $\rho(\omega)$ 
is negative-definite almost everywhere for $\omega > m$, 
except for a positive $\delta$-functional peak with the residue of $+\infty$ at $\omega = m$. 
The coexistence of negative- and positive-value regions of $\rho(\omega)$ lead to 
the anomalous increasing behavior of the effective mass $M_{\rm eff}(t)$ of gluons. 
Thus, the theoretical analysis with the Yukawa-type gluon propagator 
gives a new analytical and quantitative method for the nonperturbative gluonic phenomena.

The Yukawa function $D_{\rm Yukawa}(r) \propto e^{-mr}/r$ in the coordinate space is highly analytic,   
and this analyticity plays an important role in deriving the analytical expressions for 
$D_0(t)$, $M_{\rm eff}(t)$, and $\rho(\omega)$.
On the other hand, the momentum-space propagator $\tilde D_{\rm Yukawa}(p^2) \propto (p^2+m^2)^{-3/2}$ includes 
a singular cut stemming from the square root. 
If this singularity is taken seriously, there would arise a problem, 
because analyticity of the Green function is important in quantum field theories, 
{\it e.g.}, in the Wick rotation converting between the Euclidean space and the Minkowski space. 
Of course, this Yukawa-type function is an approximate form for the Landau-gauge gluon propagator in the region of $r=0.1 \sim 1$fm. 
There would be more regular and more thorough expression for the gluon propagator.

In this paper, we have mainly considered the gluon propagator in the coordinate space instead of the momentum space,
since the coordinate-space variable is more directly obtained in lattice QCD.
For the confined particles, however, the reason to use the momentum representation would be less clear, 
compared with ordinary particles. 
In the ordinary particles, the momentum representation of the Green function is clearly useful 
to express the pole structure, to distinguish the on-shell and off-shell states, and so on.
However, for the confined field, there is no on-shell state, {\it i.e.}, no physical asymptotic state, 
so that there is no definite reason to use the momentum representation, besides the total momentum conservation. 
Indeed, it is difficult to image the non-zero momentum-space propagator with no pole, 
and the coordinate-space representation may be more convenient for some description of the confined field,
similar to potential problems in quantum mechanics, 
where the coordinate-space wave-function is convenient.

The Yukawa-type gluon propagator $D_{\rm Yukawa}(r)$ includes $e^{-mr}$ as the main reduction factor in the infrared region, 
and $m \simeq 0.6$GeV can be regarded as the infrared effective gluon mass. 
In terms of the infrared reduction, a simple ``constituent gluon picture" 
may be approximately obtained as $M_{\rm GB} \simeq 2m$ for the glueball mass $M_{\rm GB}$.
In general, the glueball mass $M_{\rm GB}$ can be estimated from the infrared reduction of the glueball correlator  
$G_{\rm GB}(x-y)=\langle \Phi_{\rm GB}(x) \Phi_{\rm GB}(y) \rangle$ 
with the glueball operator, {\it e.g.},  $\Phi_{\rm GB} = G_{\mu\nu}^aG_{\mu\nu}^a$ for the scalar glueball, 
with the field strength tensor $G_{\mu\nu}$ \cite{R05,ISM02}.
By the Wick contraction, $G_{\rm GB}(x)$ can be expressed by some combination of the gluon Green functions. 
In the framework with the Yukawa-type gluon propagator, the leading reduction term of $G_{\rm GB}(x)$ 
can be expressed with some derivative of $\{D_{\rm Yukawa}(r)\}^2$ for large $r$. 
Then, $G_{\rm GB}(x)$ includes $e^{-2mr}$ as one of the main infrared reduction factors, 
and the lowest-glueball mass is roughly estimated as $M_{\rm GB} \simeq 2m = 1.2 \sim$ 1.3GeV, 
by neglecting all the prefactor and higher-order terms, which express the interactions between gluons. 
In spite of the crude estimate, this value gives the same order of the lowest-glueball mass of 
about 1.5GeV obtained in lattice QCD \cite{R05,ISM02}.

In this subject, there remain difficult problems related to 
the confinement mechanism, the infinite-volume limit, the Gribov copies, and the gauge dependence. 
For example, it is a highly difficult mathematical problem 
to find out the precise description of the physical hadronic states like glueballs 
in terms of the confined particles, quarks and gluons. 
As other example, Zwanziger's theorem \cite{Z91920204} is derived from the argument of the Gribov horizon:  
all connected gluon correlation functions including the gluon propagator must vanish at zero momentum 
in the infinite-volume limit.
However, our lattice QCD results and the Yukawa-type propagator $\tilde D(p^2)$ indicate 
an infrared non-vanishing property of the gluon propagator as $\tilde D(p^2=0)=4\pi^2A/m^2$.
This may be due to the absence of an infinite-volume effect in the Deep-IR region, 
as is conjectured by analytical studies \cite{Z91920204,SAH9701}.
For this problem, it is desired to clarify the Deep-IR behavior of the gluon propagator,  
and recent huge-volume lattice studies \cite{BIMPS0907,SO0607,CM0708} 
and a recent analytical study based on the Schwinge-Dyson equation \cite{K09} 
also indicate the infrared non-vanishing property of the gluon propagator.
As for the Gribov-copy problem, it is reported that the Gribov-copy effect is quantitatively rather small 
in the actual lattice QCD calculation for the Landau-gauge gluon propagator \cite{C9798,FN04},  
while the ghost propagator slightly suffers from it. 
However, this is a fundamental problem in QCD, and it would be serious in the argument of the large-volume limit, 
so that it is also desired to remove the Gribov copies and to extract the fundamental moduli region.

Finally, we discuss the Yukawa-type gluon propagation and a possible dimensional reduction 
due to the stochastic behavior of the gluon field in the infrared region. 
As shown in this paper, the Landau-gauge gluon propagator is well described 
by the Yukawa function in {\it four}-dimensional Euclidean space-time. 
However, the Yukawa function $e^{-mr}/r$ is a natural form 
in {\it three}-dimensional Euclidean space-time, 
since it is obtained by the three-dimensional Fourier transformation of 
the ordinary massive propagator $(p^2+m^2)^{-1}$. 
In fact, the Yukawa-type propagator has a ``three-dimensional" property.
In this sense, as an interesting possibility, 
we propose to interpret this Yukawa-type behavior of the gluon propagation 
as an ``effective reduction of the space-time dimension".

Such a ``dimensional reduction" sometimes occurs in stochastic systems, 
as Parisi and Sourlas pointed out 
for the spin system in a random magnetic field \cite{PS79}.
On the infrared dominant diagrams, the $D$-dimensional system 
coupled to the Gaussian-random external field 
is equivalent to the $(D-2)$-dimensional system without the external field.  
In fact, the space-time dimension of the theory is apparently reduced by two.
For the system coupled to the Gaussian-random external source, 
the dimensional reduction is universal and is associated with a hidden supersymmetry:
in the superspace formalism with $(x^\mu, \theta, \bar \theta)$, 
the integration over two Grassmann-variables ($\theta, \bar \theta$)
reduces the space-time coordinates $x^\mu$ by two \cite{PS79}.

\begin{figure}[h]
\begin{center}
\includegraphics[scale=0.48]{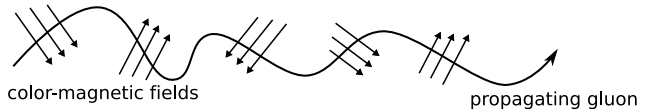}
\caption{
\label{FigGluonProgagation}
A schematic figure for a propagating gluon in the QCD vacuum. 
The QCD vacuum is filled with color-magnetic fields 
which are stochastic at an infrared scale, and 
the gluon propagates in the random color-magnetic fields.
}
\end{center}
\end{figure}

We note that the gluon propagation in the QCD vacuum resembles 
the situation of the system coupled to the stochastic external field.
In fact, as is indicated by a large positive value of the gluon condensate 
$\langle G_{\mu\nu}^aG_{\mu\nu}^a\rangle >0$ in the Minkowski space,  
the QCD vacuum is filled with a strong color-magnetic field \cite{R05,GSS07,S77,NO78AO80}, 
which can contribute spontaneous chiral-symmetry breaking \cite{ST9193}, 
and the color-magnetic field is considered to be highly random 
at an infrared scale \cite{NO78AO80,VW00,IS9900}.
Since gluons interact each other, 
the propagating gluon is violently scattered by the other gluon fields which are randomly condensed 
in the QCD vacuum at the infrared scale, as schematically shown in Fig.\ref{FigGluonProgagation}.

Actually in the infrared region, the gluon field shows a strong randomness due to the strong interaction, 
and this infrared strong randomness is considered to be responsible for color confinement, 
as is indicated in strong-coupling lattice QCD \cite{W74KS75,R05}. 
In the chiral random matrix theory for QCD \cite{VW00}, 
the infrared randomness of gluons is taken account as an essence of QCD in a simplified manner, 
and the gluon field is replaced by a Gaussian-random external field coupled to the quark field. 

Notice that there are two kinds of randomness in the gluon field: 
one is a completely-random gauge degree of freedom, which is fake, 
and the other is a net physical gluonic degree of freedom, which is not completely but highly random at the infrared scale. 
For the argument of physical randomness, these two concepts should be clearly divided, 
since the gauge degree of freedom is just a fake. 
Even after the removal of the fake gauge degree of freedom by gauge fixing, 
the gluon field exhibits a strong randomness \cite{IS9900} 
accompanying a quite large fluctuation at the infrared scale.

As a generalization of the Parisi-Sourlas mechanism, 
we conjecture that the infrared structure of a theory 
in the presence of the quasi-random external field 
in higher-dimensional space-time has a similarity to 
the theory without the external field in lower-dimensional space-time.
From this point of view, the Yukawa-type behavior of gluon propagation 
may indicate an ``effective reduction of the space-time dimension" by one,  
due to the stochastic interaction between the propagating gluon and the other gluon fields 
in the QCD vacuum, of which net physical fluctuation is highly random 
at the infrared scale.

In this paper, the Yukawa-type gluon propagator is obtained phenomenologically from lattice QCD results, 
but we expect some deeper theoretical reasons for the Yukawa-type propagation, which 
may be an effective reduction of the space-time dimension, 
due to the stochastic behavior of the infrared gluon field.
In any case, the Yukawa-type gluon  propagator would provide 
a new analytical framework for the study of nonperturbative QCD. 

\begin{acknowledgments}
H.~S. is grateful to Professor J.M.~Cornwall for his useful suggestions 
on the dynamical gluon mass. 
H.~S. is supported in part by the Grant for Scientific Research [(C) No.~19540287]
from the Ministry of Education, Culture, Science and Technology (MEXT) of Japan.
This work is supported by the Global COE Program,
``The Next Generation of Physics, Spun from Universality and Emergence" at Kyoto University.
The lattice QCD calculations have been done on NEC-SX8 at Osaka University.
\end{acknowledgments}

\appendix
\section{Fourier transformations}
In this Appendix, we derive several Fourier transformations 
used in Sec.\ref{sec:functionalforms}.

\subsection{The Yukawa function in the four-dimensional Euclidean space-time}

We first derive the four-dimensional Fourier transformation 
of the Yukawa function $e^{-mr}/r$ by calculating 
\begin{equation}
I_{\rm Yukawa} = \int d^4 x \ e^{-ip\cdot x}\frac{e^{-mr}}{r},
\label{ft:Yukawa}
\end{equation}
with $r \equiv (x_\mu x_\mu)^{1/2}$. 
We use the polar coordinate 
$(r, \theta_0,\theta_1,\theta_2)_{\rm polar}$
in four-dimensional Euclidean space-time, 
and choose the axis to satisfy 
$p\cdot x= pr \cos \theta_2$, without loss of generality. 
Then, the Fourier integral (\ref{ft:Yukawa}) is expressed as
\begin{eqnarray}
I_{\rm Yukawa}
&=& 4\pi \int_0^\pi d\theta_2 \sin^2\theta_2 \int_0^\infty dr \ r^2 e^{-(m-ip\cos\theta_2)r}
\nonumber \\
&=& 4\pi \int_0^\pi d\theta_2 \sin^2 \theta_2 \frac{1}{(m-ip\cos\theta_2)^3} \int_0^\infty
dt \ t^2 e^{-t} \nonumber \\
&=& 8\pi \int_0^\pi d\theta_2 \sin^2 \theta_2 \frac{1}{(m-ip\cos\theta_2)^3}.
\end{eqnarray}
Here, we have replaced $(m-ip\cos \theta_2)r$ by $t$, 
and changed the integration range of $t$ 
using the analytic continuation.
We rewrite $I_{\rm Yukawa}$ with partial integration as 
\begin{eqnarray}
I_{\rm Yukawa} &=& 
\frac{4\pi i}{p} \int_0^\pi d\theta \sin \theta \frac{d}{d\theta} \left[
\frac{1}{(m-ip\cos\theta)^2} \right] \nonumber \\
&=& -\frac{4\pi i}{p} \int_0^\pi d\theta \cos \theta \cdot \frac{1}{(m-ip\cos\theta)^2} \nonumber \\
&=& -\frac{4\pi}{p} \frac{d}{dp} \int_0^\pi d\theta \frac{1}{m-ip\cos\theta}
\nonumber \\
&=& -\frac{4\pi}{p} \frac{d}{dp} \frac{\pi}{(p^2+m^2)^{1/2}}=\frac{4\pi^2}{(p^2+m^2)^{3/2}}.~~~~~~
\end{eqnarray}
Here, we have used the integral formula,
\begin{equation}
\int_0^\pi d\theta\frac{1}{a+ib\cos\theta}
= \frac{\pi}{\sqrt{a^2+b^2}} 
\quad (a,b \in \mathbb{R}).
\end{equation}
Thus, we obtain the Fourier transformation of the Yukawa function as 
\begin{equation}
I_{\rm Yukawa}=
\int d^4x e^{ip\cdot x}\frac{e^{-mr}}{r} 
= \frac{4\pi^2}{(p^2 + m^2)^{3/2}},
\label{ft:Yukawa2}
\end{equation}
and its inverse Fourier transformation gives Eq.(\ref{eq:YukawaFourier}).

\subsection{The massive propagator 
in four- and two-dimensional Euclidean space-time}
We here calculate the Fourier transformation of the massive propagator $(p^2+m^2)^{-1}$.

\subsubsection{Four-dimensional Euclidean space-time}
First, we consider the Fourier integral of $(p^2+m^2)^{-1}$ 
in the four-dimensional Euclidean space-time, 
\begin{equation}
I_{\rm 4dim} = \int \frac{d^4p}{(2\pi)^4} e^{ip\cdot x}\frac{1}{p^2+m^2}.
\end{equation}
By rotating the coordinate, we set $x = (r,0,0,0)$ without loss of generality. 
Then, the integral is expressed as
\begin{eqnarray}
I_{\rm 4dim}&=& 
\int \frac{d^3p}{(2\pi)^3}
\int_{-\infty}^\infty \frac{dp_0}{2\pi} e^{ip_0 r} \frac{1}{p_0^2 + \vec{p}^2 + m^2} 
\nonumber \\
&=& 
\int \frac{d^3p}{(2\pi)^3} \frac{1}{2\sqrt{\vec{p}^2+m^2}}e^{-\sqrt{\vec{p}^2+m^2}r},
\end{eqnarray}
with $p = (p_0, \vec{p})$.
With the three-dimensional polar coordinate of $\vec{p}$, 
$I_{\rm 4dim}$ is written as  
\begin{eqnarray}
I_{\rm 4dim} &=& 
\frac{1}{4\pi^2}  \!\!
\int_0^\infty dp p^2 \  \frac{1}{\sqrt{p^2+m^2}}e^{-\sqrt{p^2+m^2}r} 
\nonumber \\
&=& \frac{1}{4\pi^2} \int_m^\infty dE \sqrt{E^2-m^2} e^{-Er} \nonumber \\
&=& \frac{1}{4\pi^2} m^2 \int_1^\infty d\epsilon \sqrt{\epsilon^2-1} e^{-\epsilon mr},
\end{eqnarray}
with $E \equiv \sqrt{p^2+m^2}$ and $\epsilon \equiv E/m$. 
Using the integral representation of the modified Bessel function,
\begin{equation}
K_1(z) = z \int_1^\infty dt e^{-zt} (t^2-1)^{1/2} \quad ({\rm Re} \ z > 0),
\label{eq:IntegFormModBessel}
\end{equation}
we obtain the Fourier transformation formula,
\begin{equation}
I_{\rm 4dim}=\int \frac{d^4p}{(2\pi)^4} e^{ip\cdot x} \frac{1}{p^2+m^2} = 
\frac{1}{4\pi^2} \frac{m}{r} K_1(mr).
\label{ft:massive4dim}
\end{equation}

\subsubsection{Two-dimensional Euclidean space-time}
Next, we consider the Fourier integral of $(p^2+m^2)^{-1}$ 
in the two-dimensional Euclidean space-time,
\begin{equation}
I_{\rm 2dim} = \int \frac{d^2p}{(2\pi)^2} e^{ip\cdot x} \frac{1}{p^2+m^2}.
\label{ft:fourier2dim}
\end{equation}
By rotating the coordinate, 
we set $x=(r,0)$ without loss of generality, 
and integrate $p_2$ as 
\begin{eqnarray}
I_{\rm 2dim} &=& 
\int_{-\infty}^\infty \frac{dp_1 }{2\pi}
\int_{-\infty}^\infty \frac{dp_2 }{2\pi}
e^{ip_1r} \frac{1}{p_1^2+p_2^2+m^2} 
\nonumber \\
&=& \int_{-\infty}^{\infty} \frac{dp_1}{4\pi} \frac{e^{ip_1 r}}{\sqrt{p_1^2+m^2}}
=\int_0^\infty \frac{dp_1}{2\pi} \frac{\cos(p_1 r)}{\sqrt{p_1^2+m^2}}.~~~~~~~~
\end{eqnarray}
Using Mehler's integral representation of the modified Bessel function,
\begin{equation}
K_0(z) = \int_0^\infty dt \frac{\cos(zt)}{(t^2+1)^{1/2}},
\end{equation}
we obtain the Fourier transformation formula, 
\begin{equation}
I_{\rm 2dim}=\int \frac{d^2p}{(2\pi)^2} e^{ip\cdot x}\frac{1}{p^2+m^2} = \frac{1}{2\pi} K_0(mr).
\end{equation}

\subsection{The dipole-type propagator}
We deal with the Fourier integral of the dipole-type propagator 
$(p^2+m^2)^{-2}$ in the four-dimensional Euclidean space-time,
\begin{eqnarray}
I_{\rm dipole}
&=& \int \frac{d^4p}{(2\pi)^4} e^{ip\cdot x} \frac{1}{(p^2+m^2)^2} \nonumber \\
&=& - \frac{1}{2m} \frac{d}{dm} 
\left[ \int \frac{d^4p}{(2\pi)^4} e^{ip\cdot x} \frac{1}{p^2+m^2} \right].~~~~~
\end{eqnarray}
Using the Fourier transformation (\ref{ft:massive4dim}), we get 
\begin{eqnarray}
I_{\rm dipole}&=& 
- \frac{1}{2m} \frac{d}{dm} 
\left[ \frac{1}{4\pi^2} \frac{m}{r} K_1(mr) \right] \nonumber \\
&=& - \frac{1}{8\pi^2} \left[ \frac{1}{mr} K_1(mr) + \frac{1}{r} \frac{d}{dm} K_1(mr) \right].~~~~~~
\end{eqnarray}
From the relation of the modified Bessel function,
\begin{equation}
z K_\nu^\prime (z) + \nu K_\nu(z) = - zK_{\nu-1}(z),
\end{equation}
we obtain the Fourier transformation formula, 
\begin{equation}
I_{\rm dipole}=\int \frac{d^4p}{(2\pi)^4} e^{ip\cdot x} \frac{1}{(p^2+m^2)^2} = \frac{1}{8\pi^2}K_0(mr).
\end{equation}

\section{Deep-IR corrected gluon propagator}

In the Deep-IR region, 
there has been reported to appear 
some deviation on the momentum-space gluon propagator 
$\tilde D(p^2)$ between small-size lattice data 
and huge-volume lattice data \cite{BIMPS0907,SO0607,CM0708}.  
In this Appendix, we demonstrate that the deviation 
in the Deep-IR region does not affect the Yukawa-type behavior of 
the coordinate-space gluon propagator $D(r)$ 
in the IR/IM region of $r=0.1 \sim 1.0$fm.

\begin{figure}[h]
\begin{center}
\includegraphics[scale=1.1]{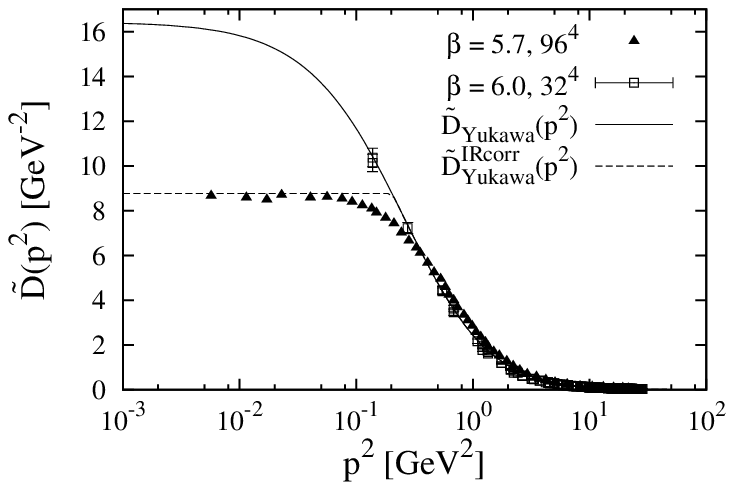}
\caption{\label{fig:hugeLattice} 
The infrared behavior of the gluon propagator $\tilde D(p^2)$. 
The triangle symbols denote recent 
huge-volume lattice data taken from Ref.\cite{BIMPS0907}.
The solid line denotes the Yukawa-type propagator 
$\tilde{D}_{\rm Yukawa}(p^2)$, and
the dashed line the Deep-IR-corrected propagator 
$\tilde{D}_{\rm Yukawa}^{\rm IRcorr}(p^2)$ with 
$p_{\rm IR} = 0.45$GeV.
}
\end{center}
\end{figure}

Figure \ref{fig:hugeLattice} shows the scalar-type gluon propagator 
$\tilde{D}(p^2)$ in the recent lattice-QCD calculation 
with a huge volume $96^4$ at $\beta=5.7$, 
taken from Ref.\cite{BIMPS0907}.
Here, a renormalization constant is multiplied 
for the huge-volume lattice data 
so as to adjust them to the renormalization condition 
(\ref{RenormCond}) at $\mu$=4GeV.
In the momentum space, the true gluon propagator 
$\tilde D(p^2)$ turns out to take a saturated value 
smaller than the Yukawa-type propagator 
$\tilde D_{\rm Yukawa}(p^2)$ 
in the Deep-IR region of $p < 0.5$GeV.
In other words, $p \simeq 0.5$GeV is the lower bound on 
the applicability of the Yukawa-type propagator $\tilde D_{\rm Yukawa}(p^2)$ 
to the gluon propagator.

Taking account of the deviation in the Deep-IR region, 
we define the Deep-IR-corrected momentum-space propagator as
\begin{eqnarray}
\tilde{D}_{\rm Yukawa}^{\rm IRcorr}(p^2) =
\begin{cases}
\tilde{D}_{\rm Yukawa}(p^2)  &p \geq p_{\rm IR}  \\
\tilde{D}_{\rm Yukawa}(p_{\rm IR}^2) ~ {\rm (const.)}  &p \leq p_{\rm IR} \\
\end{cases}
\end{eqnarray}
with the IR-saturation momentum of $p_{\rm IR} = 0.45{\rm GeV}$.
This value of $p_{\rm IR}$ is determined 
so as to consist with the huge-volume lattice result in the Deep-IR region.
Using this Deep-IR-corrected propagator 
$\tilde{D}_{\rm Yukawa}^{\rm IRcorr}(p_{\rm IR}^2)$, 
we calculate the Deep-IR-corrected coordinate-space propagator
by the Fourier transformation as 
\begin{equation}
D_{\rm Yukawa}^{\rm IRcorr}(r) = \int \frac{d^4p}{(2\pi)^4} 
e^{-ip\cdot x}\tilde{D}_{\rm Yukawa}^{\rm IRcorr}(p^2).
\end{equation}

\begin{figure}[h]
\begin{center}
\includegraphics[scale=1.1]{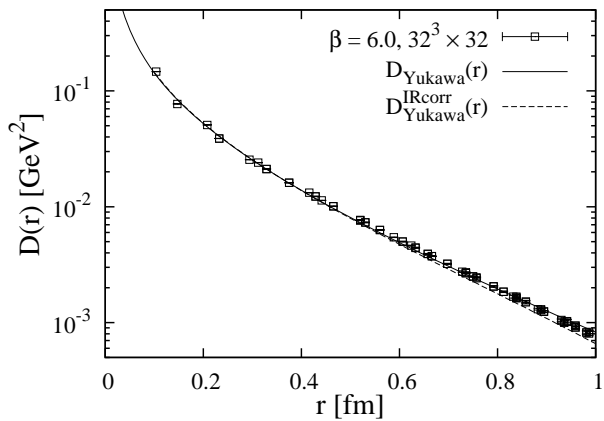}
\caption{\label{fig:hugeLatticeCorr}
The Yukawa-type propagator $D_{\rm Yukawa}(r)$ (solid line),
and Deep-IR-corrected propagator $D_{\rm Yukawa}^{\rm IRcorr}(r)$ 
(dashed-line), together with the lattice data. 
The difference between them 
is fairly small in the IR/IM region of $r=0.1 \sim 1.0$fm.
}
\end{center}
\end{figure}

In Fig.\ref{fig:hugeLatticeCorr}, 
we show the Yukawa-type propagator $D_{\rm Yukawa}(r)$
and this Deep-IR-corrected propagator $D_{\rm Yukawa}^{\rm IRcorr}(r)$, 
together with the lattice QCD data. 
The difference between $D_{\rm Yukawa}(r)$ 
and $D_{\rm Yukawa}^{\rm IRcorr}(r)$ is fairly small for $r = 0.1 \sim 1.0$ fm.

Thus, in the coordinate space, 
the Yukawa-type function $D_{\rm Yukawa}(r)$ 
works well for the IR/IM region of $r = 0.1 \sim 1.0$ fm, 
even after the correction in the Deep-IR region.

\end{document}